\documentclass[10pt,twocolumn,twoside]{IEEEtran}
\usepackage{float}
\usepackage{graphicx}
\usepackage{latexsym}
\usepackage{amsmath}
\usepackage{amssymb}
\usepackage{epsfig}
\usepackage{cite}
\usepackage{algorithmic}
\usepackage{overpic}
\usepackage{multirow}
\usepackage[svgnames]{xcolor}
 \usepackage{empheq}
\usepackage{colortbl,array}
\usepackage{multirow,bigdelim}
\usepackage{subfigure}
\usepackage{amsthm}
\usepackage{booktabs,amsmath}
\usepackage{fancyhdr, graphicx, amsmath, amssymb}
\usepackage[linesnumbered,ruled]{algorithm2e}
\usepackage{arydshln}
\usepackage[english]{babel}
\usepackage{xcolor}
\usepackage{tcolorbox}
\usepackage{soul}
\usepackage[hidelinks]{hyperref}
\usepackage[framemethod=default]{mdframed}
\usepackage{hyperref}
\usepackage{mathtools}
\usepackage{booktabs}
\usepackage[export]{adjustbox}
\usepackage{tabularx}
\usepackage{makecell}
\usepackage[customcolors]{hf-tikz}
\usepackage{nicematrix}
\usepackage{tikz}
\usepackage{wrapfig}

\newtheorem{thm}{Theorem}
\newtheorem{lem}{Lemma}
\newtheorem{rem}{Remark}

\newtheorem{asm}{Assumption}
\newtheorem{defa}{Definition}

\newtheorem{prom}{Problem}

\newcommand{\red}[1]{\textcolor{red}{#1}}

\ifCLASSINFOpdf
\else
\fi

\begin{document}

\title{Cost Function Learning in Memorized Social Networks with Cognitive Behavioral Asymmetry}

\author{Yanbing~Mao, Jinning~Li, Naira~Hovakimyan, Tarek~Abdelzaher, and Christian~Lebiere
\thanks{Y.~Mao is with the Engineering Technology Division, Wayne State University, Detroit, MI 48201, USA (e-mail: hm9062@wayne.edu).}
\thanks{J.~Li is with the Department of Computer Science, University of Illinois at Urbana-Champaign, Urbana, IL 61801, USA (e-mail: jinning4@illinois.edu).}
\thanks{N.~Hovakimyan is with the Department of Mechanical Science and Engineering, University of Illinois at Urbana--Champaign, Urbana, IL 61801, USA (e-mail: nhovakim@illinois.edu).}
\thanks{T.~Abdelzaher is with the Department of Computer Science, University of Illinois at Urbana-Champaign, Urbana, IL 61801, USA (e-mail: zaher@illinois.edu).}
\thanks{C.~Lebiere is with the Department of Psychology, Carnegie Mellon University, Pittsburgh, PA 15213, USA (e-mail: cl@cmu.edu).}
\thanks{This work was supported in part by DOD HQ00342110002 and DARPA W911NF-17-C-0099.}}

\maketitle
\begin{abstract}
This paper investigates the cost function learning in social information networks, wherein the human memory and cognitive bias are explicitly taken into account.  We first propose a model for social information-diffusion dynamics, with a focus on systematic modeling of asymmetric cognitive bias represented by confirmation bias and novelty bias. Building on the dynamics model, we then propose the M$^{3}$IRL -- a Memorized Model and Maximum-entropy based Inverse Reinforcement Learning -- for learning cost functions. Compared with the existing model-free IRLs, the characteristics of M$^{3}$IRL are significantly different here: no dependency on the Markov Decision Process principle, the need of only a single finite-time trajectory sample, and bounded decision variables. Finally, the effectiveness of the proposed social information-diffusion model and the M$^{3}$IRL algorithm are validated by the online social media data.
\end{abstract}

\IEEEpeerreviewmaketitle

\section{Introduction}
\IEEEPARstart{W}{ith} the holistic combination of artificial intelligence, communication, and information technology, social media has pushed our society into far-reaching and globally connected communities, wherein people can express a wide variety of unfiltered opinions, sentiments, and emotions, regarding, e.g., health, illness, and health services \cite{chou2009social}. The consequential challenges we are facing are the widespread low-quality information (e.g., misinformation and disinformation) from malicious information sources with inauthentic behavior \cite{wojtowicz2020addressing,gollust2020emergence,swire2019public}. To fight malicious information sources, the game-theoretic framework of competitive information diffusion was initially formulated in \cite{mao2019competitive,mao2019impact,dhamal2018optimal,xx1,xx2,xx3}, where the malicious information sources disperse the low-quality information while the defenders spread the truthful information to counter the influence of low-quality information on public. The game-theoretic formulations  rely  on  a common assumption that the game players know each other's cost function for decision-making. To pave the way to fight against malicious information sources in a game-theoretic adversarial setting, we thus investigate the problem: \emph{How to learn the implicit cost functions of game players?}

Cost function learning is a much more challenging and deeper problem, which has deep roots in the inverse optimal control (IOC) \cite{kalman1964linear} and then later studied in the context of inverse reinforcement learning (IRL) \cite{choi2012nonparametric, 2012-cioc, boularias2011relative, ziebart2008maximum, ziebart2010modeling}.  However,  IRLs in social information networks with humans in the loop are not explored yet. The practical challenges hindering the exploration include:\\
-- Human Memory: Recent investigation on social (mis)-information spread from the perspective of cognitive architecture reveals the significant influence of human memory in the information consumption \cite{cck,stanley2016comparing}, which contradicts with the common assumption of Markov Decision Process (MDP) in both IOC and IRL \cite{choi2012nonparametric, 2012-cioc, boularias2011relative, ziebart2008maximum, ziebart2010modeling,jin2019inverse,schwarting2019social, abbeel2004apprenticeship}.\\
-- Dynamic External Stimulus: The dynamic external stimulus pushes the cost functions of decision-making to be dynamic accordingly. For example, before the outbreak of COVID-19 crisis in the US, the news agency Fox News dispersed the misinformation that the COVID-19 is a hoax, which is driven by the event of US President impeachment, and lately changed the tune due to the possible lawsuit  \cite{fox}. In the dynamic scenarios,  the most recent trajectory of evolving opinions can yield a much more accurate inference of cost functions than a large number of trajectory samples collected under different external stimulus.

The seminal model-based IRL with local optimality \cite{2012-cioc} provides a potential building block to address the challenges induced by human memory and dynamic external stimulus. Specifically, the model-based IRL needs only a single finite-time trajectory and removes the assumption that the expert demonstrations are globally probabilistically optimal. Inspired by the model-based IRL, we propose the M$^{3}$IRL: a Memorized Model and Maximum-entropy based Inverse Reinforcement Learning with local optimality.

The proposed M$^{3}$IRL applying in social information-diffusion networks is model\red{-}based. The inference accuracy of cost functions via M$^{3}$IRL thus relies on the trustworthiness of social information-diffusion model. Therefore, the challenge moving forward is \textit{What is a social information-diffusion aggregate model that can well describe how humans consume and propagate information as well as how beliefs evolve?} Recently, with the wide use of social media \cite{xu2020paradox}, in conjunction with automated news generation with the help of artificial intelligence technologies \cite{giridhar2019social,cui2019semi}, the dynamics of social information-diffusion has gained vital importance in studying misinformation spread and political polarization. Meanwhile, it has been revealed that the cognitive bias, especially the confirmation bias and the novelty bias play a key role in the misinformation spread and polarization evolution.  In particular,  it is well understood that the confirmation bias helps create ``echo chambers" within online social networks \cite{lazer2018science, vicario2019polarization}, in which misinformation and polarization thrive \cite{del2016spreading, cinelli2020echo}. Recently, Abdelzaher et al. in \cite{abdelzaher2020paradox} and Xu et al. in \cite{xu2020paradox} revealed the significant influence of consumer preferences for outlying content (due to novelty bias) on the opinion polarization in the modern era of information overload. Hence, the challenge pertaining to modeling the information diffusion dynamics is \emph{how to capture human cognitive bias in information consumption and spread?}

The Hegselmann-Krause (HK) model \cite{hegselmann2002opinion} has been recognized for the capability of capturing confirmation bias \cite{del2017modeling}, through imposing a bounded confidence on opinion distances. The HK model involves a discontinuity in the influence impact, i.e., an individual completely ignores the opinions that are ``too far" from hers. The discontinuity however renders the steady-state analysis difficult. As a remedy, the tractable continuous (state-dependent) social-influence models were proposed in \cite{jabin2014clustering,motsch2014heterophilious,mao2019impact,mao2020social,mao2020inference} to capture the confirmation bias. We note that in the most of the social problems, e.g., president election and product rating, humans hold the asymmetric cognitive bias (i.e., the same opinion distance can result in different influence weights) in their opinion evolutions. However, the HK model with symmetric confidence boundary \cite{hegselmann2002opinion} and the continuous models \cite{jabin2014clustering,motsch2014heterophilious,mao2019impact,mao2020social,mao2020inference} can only capture the symmetric confirmation bias (i.e., the same opinion distance results in same influence weights), and the HK model with asymmetric confidence boundaries \cite{hegselmann2002opinion} can only partially capture the asymmetric confirmation bias. Additionally, the existing information diffusion models do not consider capturing the novelty bias yet \cite{friedkin1990social,degroot1974reaching,hegselmann2002opinion,jabin2014clustering,motsch2014heterophilious,mao2019impact,mao2020social,mao2020inference}. To obtain a fairly accurate model of social information for the proposed M$^{3}$IRL, we provide a systematic modeling guidance of asymmetric cognitive bias, represented by confirmation bias and novelty bias.

The contributions of this paper are summarized as follows.
\begin{itemize}
  \item We propose a model of social information-diffusion dynamics, which explicitly takes subconscious bias, human memory, confirmation bias and novelty bias into account. Meanwhile, we provide the systematic modeling guidance for capturing asymmetric confirmation bias and asymmetric novelty bias.
  \item Building on the proposed model of social information diffusion, we propose the M$^{3}$IRL for learning the cost functions of target individuals in social networks with humans in the loop.
  \item Given a library of basis functions that constitute the cost functions in social information networks, we validate the effectiveness of proposed M$^{3}$IRL using the online social media data.
\end{itemize}

\section{Preliminaries}
\subsection{Notation}
The social system is composed of $n$ individuals. The interaction among individuals is modeled by a digraph $\mathfrak{G} = (\mathbb{V}, \mathbb{E})$, where $\mathbb{V}$ = $\left\{\mathrm{v}_{1}, \ldots,  \mathrm{v}_{n}\right\}$ is the set of vertices representing the individuals, and $\mathbb{E} \subseteq  \mathbb{V} \times \mathbb{V}$ is the set of edges of the digraph $\mathfrak{G}$  representing the influence structure. Other notations that are used throughout this paper are included in Table \ref{notation}. 
\begin{table}[ht] \footnotesize{
\centering
\caption{Table of Notation}
\begin{tabular}{|l|}
\hline
$\mathbb{R}^{n}$:~set of $\emph{n}$-dimensional real vectors  \\ \hline
$\mathbb{R}^{m \times n}$:~set of  $m \times n-$dimensional real matrices  \\ \hline
 $[W]_{i,j}$:~ element at row $i$ and column $j$ of matrix $W$ \\ \hline
$\left|  \cdot  \right|$:~ cardinality (i.e., size) of a set,  matrix determinant, or  absolute value\\ \hline
$\mathbf{O}$:~  zero matrix with compatible dimensions  \\ \hline
$\top$:~transposition of a matrix or vector  \\ \hline
 $[x~;~ y]:= [x^\top, ~y^\top ]^\top$ \\ \hline
${\underline{\mathfrak{g}}_g}(\cdot)$: $\mathbb{R} \rightarrow \mathbb{R}_{\ge 0}$, ~~${\overline{\mathfrak{g}}_g}(\cdot)$: $\mathbb{R} \rightarrow \mathbb{R}_{\ge 0}$, ~~${\underline{\mathfrak{f}}_g}(\cdot)$: $\mathbb{R} \rightarrow \mathbb{R}$, ~~${\overline{\mathfrak{f}}_g}(\cdot)$: $\mathbb{R} \rightarrow \mathbb{R}$ \\ \hline
${\breve{c}_q}(\tilde{x}(j),u(j))$: ~cost function of individual $q$ at time $j$\\ \hline
\end{tabular}\label{notation}}
\end{table}

To end this subsection, we introduce the definitions of interested cognitive bias in particular. 
\begin{defa} 
The confirmation bias is generally referred to the cognitive behavior where a person gives larger weight to evidence that confirms his belief and undervalues evidence that could disprove it \cite{CCBB}. 
\end{defa}
\begin{defa} 
The novelty bias refers to humans' preferences for outlying content \cite{abdelzaher2020paradox}. 
\end{defa}
\begin{defa}  
The subconscious bias refers to an individual's innate opinion, which is based on inherent personal characteristics, e.g., socio-economic conditions where the individual grew up and/or lives in \cite{friedkin1990social}. 
\end{defa}

\vspace{-0.40cm}
\subsection{Social Information-Diffusion Dynamics}
\vspace{-0.10cm}
We consider the following model of social information diffusion (adopted
from our prior model \cite{mao2018spread,mao2019impact,mao2018evolution}), which will be used to derive the model-based cost function learning algorithm.
\begin{align}
\!\!\!{x_i}(k + 1) = {\alpha _i}(x,k,\tau_{i}){s_i} + \sum\limits_{j \in \mathbb{V}} {{c_{ij}}(x,k,\tau_{i}){x_j}(k)},~i \!\in\! \mathbb{V}\label{kka}
\end{align}
where we clarify the notations and variables:
\begin{itemize}
  \item ${x_i}\!\left( {k} \right) \in [-1,1]$ is individual $\mathrm{v}_{i}$'s evolving opinion at time $k$, ${s_i} \in [-1,1]$ is her subconscious bias.
  \item ${c_{ij}}(x,k,\tau_{i})$ represents the influence weight of individual $\mathrm{v}_{j}$ on  $\mathrm{v}_{i}$, and
  \begin{equation}
	{c_{ij}}(x,k,\tau_{i}) = \begin{cases}
		> 0, & \text{if } (\mathrm{v}_i,\mathrm{v}_j) \in \mathbb{E}\\
		= 0, & \text{otherwise}
	\end{cases} \label{fil}
  \end{equation}
where the $\tau_{i}$ denotes individual i's memory horizon. For example, at current time $k$, indivial $\mathrm{v}_{i}$ has a memory of a topic information over the time $k -1$, $k -2$, ..., $k - \tau_{i}$. 
  \item The state-dependent influence weight ${c_{ij}}(x,k,\tau_{i})$ is proposed to capture $\mathrm{v}_{i}$'s cognitive bias induced by the conjunctive confirmation bias and novelty bias:
 \begin{align}
  {c_{ij}}(x,k,\tau_{i}) &= \overline{c}( {{{\overline{\mathrm{x}}}_i}(x,k,\tau_{i}),{x_j}(k)}) \nonumber\\
 &\hspace{2.10cm} +  \underline{c}( {{{\underline{\mathrm{x}}}_i}(x,k,\tau_{i}),{x_j}(k)}), \label{sdw2b}
  \end{align}
  where $\overline{c}( {{{\overline{\mathrm{x}}}_i}(x,k,\tau_{i}),{x_j}(k)}) \geq 0$ is proposed to capture the novelty bias, and $\underline{c}( {{{\underline{\mathrm{x}}}_i}(x,k,\tau_{i}),{x_j}(k)}) \geq 0$  describes the confirmation bias.
  \item The ${{\overline{\mathrm{x}}}_i}(x,k,\tau_{i})$ in \eqref{sdw2b} denotes individual $\mathrm{v}_{i}$'s sensed expectation from her memory of surrounding opinions over the memory horizon $\left\{ {k \!-\! \tau_{i} ,k \!-\! \tau_{i}  \!+\! 1, \ldots ,k} \right\}$. The surroundings in real life can include individual's neighbors and the information sources she follows. The ${{\overline{\mathrm{x}}}_i}(x,k,\tau_{i})$ is defined as
\begin{align}
{{\overline{\mathrm{x}}}_i}(x,k,\tau_{i}) \triangleq \frac{{\sum\limits_{t = k - {\tau _i}}^k {{m_i}(t) \cdot \sum\limits_{j = 1}^{\left| \mathbb{V} \right|} {{c_{ij}}(x,t,{\tau _i}){x_j}(t)} } }}{{\sum\limits_{v = k - {\tau _i}}^k {\sum\limits_{p = 1}^{\left| \mathbb{V} \right|} {{m_i}(v) \cdot {c_{ip}}(x,v,{\tau _i})} } }}. \label{sdw2}
\end{align}
The time-varying function ${m_i}(t)$ in Eq. \eqref{sdw2} is proposed to indicate the influence of memory horizon on the sensed expectation, which satisfies
\begin{align}
0 \leq {m_i}(t) \leq {m_i}(t+1), ~~~\forall t \in \mathbb{N}. \label{sdw2aa}
\end{align}

\item The ${{\underline{\mathrm{x}}}_i}(x,k,\tau_{i})$ in Eq. \eqref{sdw2b} denotes individual $\mathrm{v}_{i}$'s sensed expectation from her own memory:
\begin{align}
{{\underline{\mathrm{x}}}_i}(x,k,\tau_{i}) \triangleq \frac{{\sum\limits_{t = k - {\tau _i}}^k {{m_i}\left( t \right) \cdot {x_i}\left( t \right)} }}{{\sum\limits_{v = k - {\tau _i}}^k {{m_i}\left( v \right)} }}. \label{sdw2kk}
\end{align}
\item  $\alpha_{i}(x,k,\tau_{i}) \geq 0$  is the ``resistance parameter'' of individual $\mathrm{v}_{\mathrm{i}}$ on her subconscious bias. To guarantee $x_{i}(k) \in [-1, 1]$ for $\forall k \in \mathbb{N}$ and $\forall i \in \mathbb{V}$, it is determined in such a way that
\begin{align}
\!\!{\alpha_{i}}(x,k,\tau_{i}) + \sum\limits_{j \in \mathbb{V}} {{c_{ij}}(x,k,\tau_{i})}  = 1, ~~\forall i \in \mathbb{V}.\label{sdw3}
\end{align}
\end{itemize}

\begin{rem}
The imposed condition Eq. \eqref{sdw2aa} indicates the decaying influence of memory horizon on the individuals' real-time sensed exceptions Eq. \eqref{sdw2} and Eq. \eqref{sdw2kk}. The ${m_i}(t)$ can be a function of decaying activation, e.g., base-level activation, proposed in the cognitive architecture \cite{stanley2016comparing,cck}.
\end{rem}

\vspace{-0.40cm}
\subsection{Problem Formulation}
\vspace{-0.10cm}
It is well understood and has been demonstrated that confirmation bias helps create ``echo chambers" within online social networks \cite{lazer2018science, vicario2019polarization}, in which misinformation and polarization thrive \cite{del2016spreading, cinelli2020echo}. Meanwhile, Abdelzaher et al. in \cite{abdelzaher2020paradox} and Xu et al. in \cite{xu2020paradox} recently revealed the significant influence of novelty bias on the opinion polarization. In addition, more than 40 years of studies in cognitive and social psychology have revealed that the asymmetry effect/bias (i.e., the distance from X to Y may be estimated differently from Y to X) is a universal phenomenon, ranging from psychological similarity estimations \cite{codol1989asymmetry} to social perception \cite{TK77}.  From the perspective of modeling the social information-diffusion dynamics, how to systemically capture the realistic asymmetric cognitive bias is not explored yet. The first problem we will address is pertaining to the model, whose solution will constitute the base of model-based cost function learning. 
\begin{prom}
What is the systematic modeling guidance for the social information-diffusion dynamics that can capture the asymmetric confirmation bias and the asymmetric novelty bias? \label{pr1}
\end{prom}

Generally, an individual has an implicit individual/joint cost function for (e.g., political-gain driven, profit-driven and curiosity-driven) decision making. Moreover, the game-theoretic formulations for fighting against malicious information sources rely on an assumption that the players know each other's cost function for decision-making \cite{mao2019impact,dhamal2018optimal}. Learning the cost functions is thus indispensable for the feasibility of game-theoretic defense strategies. Meanwhile, we note that the dynamic external stimulus pushes the cost functions of decision makers to be dynamic as well \cite{fox}. In the dynamic scenarios, the most recent trajectory rather than a larger number of trajectory samples collected under different stimulus is more desired for a more accurate inference of cost function. Building on the answer to the Problem \ref{pr1}, the accurate cost function learning constitutes the second problem.
\begin{prom}
Given the model of social information-diffusion dynamics \eqref{kka} that well captures asymmetric confirmation bias and asymmetric novelty bias, how to leverage the most recent trajectory to learn the cost functions of target individuals?  \label{pr2}
\end{prom}

\begin{figure}
\centering
\includegraphics[scale=0.38]{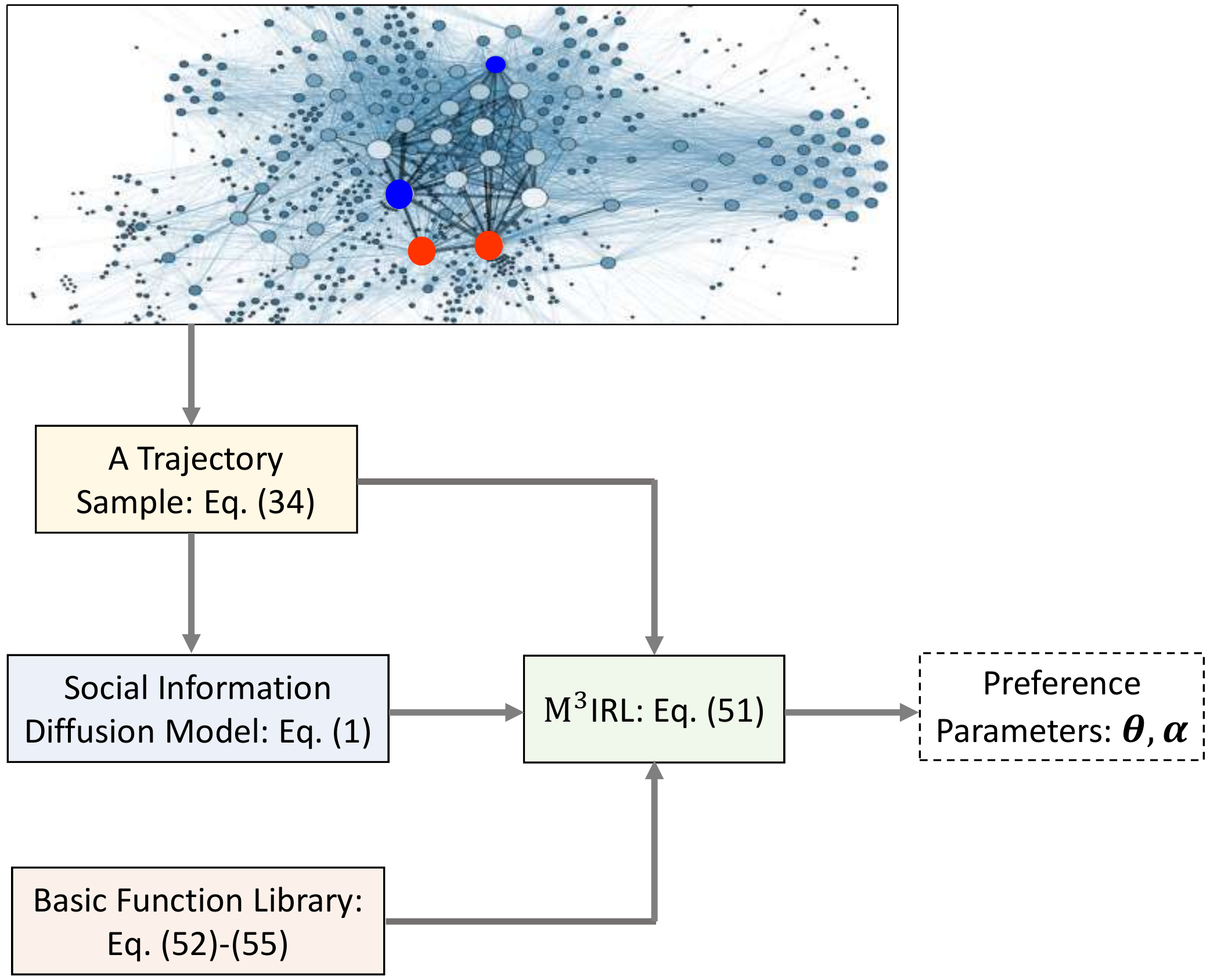}
\caption{Model-based framework of cost function learning.}
\label{cf}
\end{figure}
With the solutions to Problems \ref{pr1} and \ref{pr2}, the proposed framework of cost function learning is presented in Figure \ref{cf}, where the red and blue nodes denote the observed target individuals. Specifically, we first collect a finite-time trajectory of evolving opinions and actions. The trajectory sample is then used to fit the proposed model of social information diffusion. We next input the fitted model, the trajectory sample, and the library of basis functions to the M$^3$IRL for computing the vectors of preference parameters. We finally obtain the cost functions, which are linear combinations of basis functions. In the formulas of cost functions, the preference parameters serve as the combination coefficients associated with basis functions.

\vspace{-0.15cm}
\section{Problem \ref{pr1}: Asymmetric Cognitive Bias}
\begin{figure}
\centering
\includegraphics[scale=0.27]{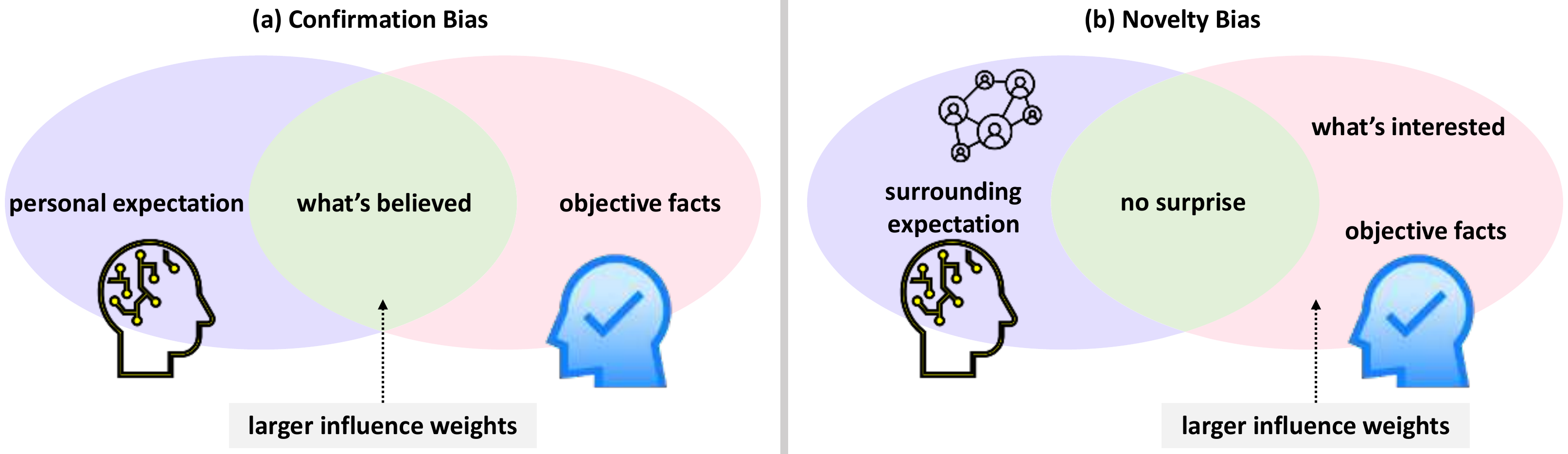}
\caption{Modeling mechanism aims at capturing the cognitive behavior due to confirmation bias and novelty bias.}
\label{mmm}
\end{figure}
We first use the confirmation bias as an example to describe the asymmetry bias phenomenon in opinion evolution, and the expected behavior that the influence weight $\underline{c}( {{{\underline{\mathrm{x}}}_i}(x,k,\tau_{i}),{x_j}(k)})$ should capture. We then extend the modeling mechanism to the novelty bias. The cognitive behavior due to confirmation bias and novelty bias that the model shall capture is described by Figure \ref{mmm}. For the sake of simplifying the presentation, we refer $x_{a}$, $x_{b}$ and ${{\underline{\mathrm{x}}}_g}$ (dropping out $x$, $k$ and $\tau_{i}$ without loss of generality) to the opinions of Alex, Bob and George (sensed from his memory), respectively. We suppose the topic being discussed is ``\emph{COVID-19 Is a Hoax}." The hierarchy representations of $x_{i}(t)$ is illustrated by Figure \ref{exep}-(i), where $1$ and $-1$ correspond to `completely opposing' and `completely supporting' the claim, respectively.

\vspace{-0.30cm}
\subsection{Behavior Due to Asymmetric Confirmation Bias}
\begin{figure}
\centering
\includegraphics[scale=0.31]{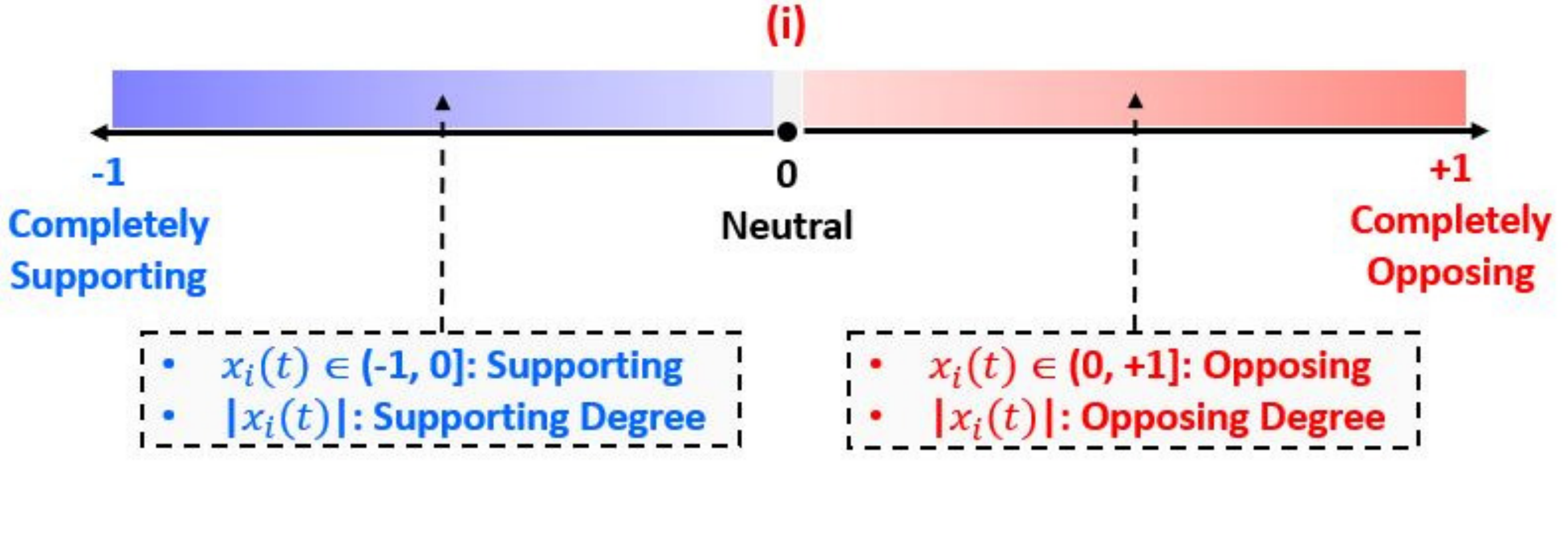}\\
\includegraphics[scale=0.34]{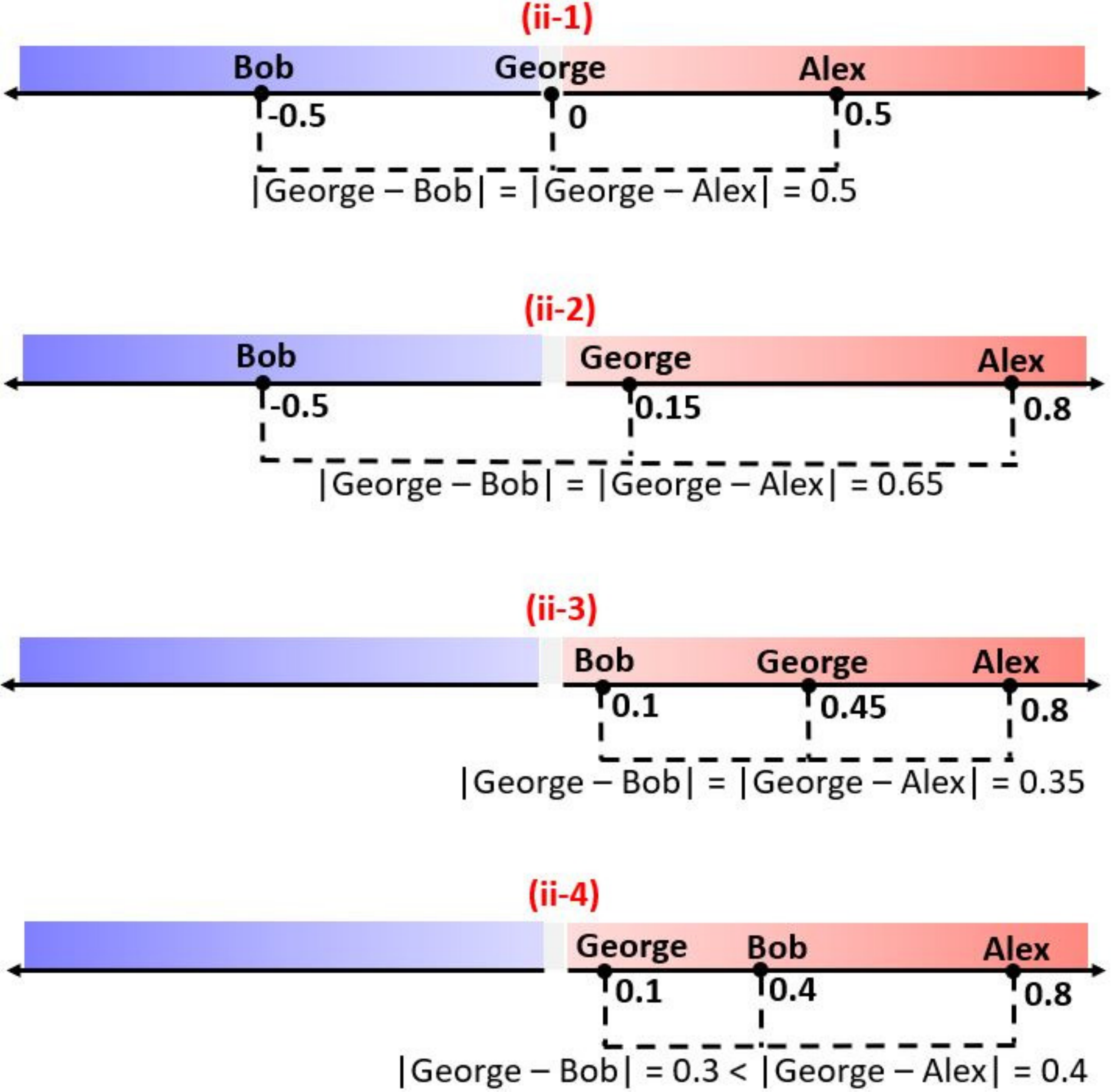}\\
\includegraphics[scale=0.34]{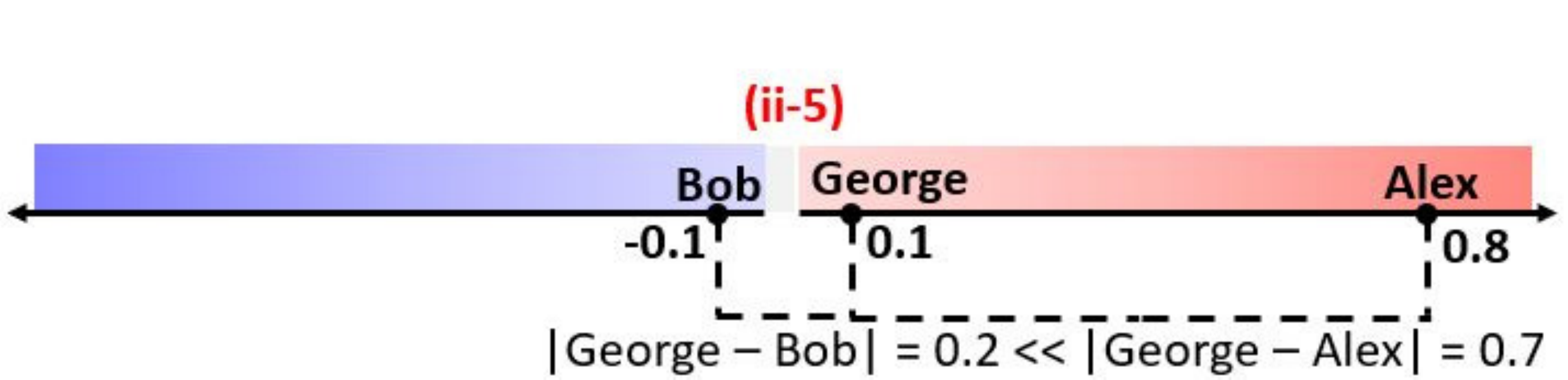}
\caption{Asymmetric cognitive bias explanation.}
\label{exep}
\end{figure}

We describe George's five behavior in reality if he holds confirmation bias towards the opinions of his neighbors Alex and Bob. The five corresponding behavioral scenarios are shown in Figure \ref{exep}-(ii-1)-(ii-5), respectively.

\underline{\emph{Figure \ref{exep}-(ii-1): Neutral Opinion with the Same Distance:}} The  ${{\underline{\mathrm{x}}}_g}  = 0$ indicates that George neither supports nor opposes the claim. In this scenario, George should treat the opinions of Alex and Bob equally, as long as they have the same distance from his opinion. This behavior is formally described by
\begin{align}
\underline{c}({{\underline{\mathrm{x}}}_g},x_{a}) = \underline{c}({{\underline{\mathrm{x}}}_g},x_{b}), \hspace{0.14cm}~\text{if}~x_{a} = -{{x}_{b}} ~\text{and}~ {{\underline{\mathrm{x}}}_g}  = 0.\label{confim4}
\end{align}

\underline{\emph{Figure \ref{exep}-(ii-2): Same Distance \& Crossing Domains:}} The example \{$x_{b} = -0.5$, ${{\underline{\mathrm{x}}}_g}  = 0.15$, $x_{a} = 0.8$\} means that
both  Alex and Bob have the same opinion distance from George's, but George and Alex are in the same domain of opposing the claim while Bob is in the other domain of supporting. In this scenario, George should favor Alex's opinion more. This behavior is formally described by
\begin{align}
&\underline{c}({{\underline{\mathrm{x}}}_g},x_{a}) > \underline{c}({{\underline{\mathrm{x}}}_g},x_{b}), \hspace{0.14cm}~\text{if}~| {{{\underline{\mathrm{x}}}_g} - x_{a}} | = | {{\underline{\mathrm{x}}}_g} - x_{b}| ~\text{and} \nonumber\\
&\hspace{4.90cm}
{{\underline{\mathrm{x}}}_g} \cdot x_{a} > 0 \geq {{\underline{\mathrm{x}}}_g} \cdot x_{b}. \label{confim3}
\end{align}

\underline{\emph{Figure \ref{exep}-(ii-3): Same Distance \& Same Domain:}} The case \{$x_{b} = 0.1$, ${{\underline{\mathrm{x}}}_g}  = 0.45$, $x_{a} = 0.8$\} indicates that
Alex and Bob have the same opinion distance from George's and they three are in the same domain of opposing the claim. But Bob is more hesitating in his opinion and more likely to leave the current domain in his next opinion evolution process, while George and Alex are more stubborn. In this scenario, George should also favor Alex's opinion more, which is described by
\begin{align}
&\underline{c}({{\underline{\mathrm{x}}}_g},x_{a}) > \underline{c}({{\underline{\mathrm{x}}}_g},x_{b}), \hspace{0.14cm}~\text{if}~| {{{\underline{\mathrm{x}}}_g} - x_{a}} | = | {{\underline{\mathrm{x}}}_g} - x_{b}| ~\text{and} \nonumber\\
&\hspace{4.75cm}
{{\underline{\mathrm{x}}}_g} \cdot x_{a} > {{\underline{\mathrm{x}}}_g} \cdot x_{b} \geq 0. \label{confim2}
\end{align}

\underline{\emph{Figure \ref{exep}-(ii-4): Same Domain \& Different Distances:}} The example \{$x_{b} = 0.4$, ${{\underline{\mathrm{x}}}_g}  = 0.1$, $x_{a} = 0.8$\} means that
1) George, Alex and Bob are in the same domain of opposing the claim, 2) both Alex and Bob are stubborn in their opinions, but 3) Bob's opposing degree is closer to George's. In this scenario, George should favor Bob's opinion more.  This expected behavior is described as
\begin{align}
&\underline{c}({{\underline{\mathrm{x}}}_g},x_{b}) > \underline{c}({{\underline{\mathrm{x}}}_g},x_{a}), \hspace{0.14cm}~\text{if}~| {{{\underline{\mathrm{x}}}_g} - x_{b}} | < | {{\underline{\mathrm{x}}}_g} - x_{a}| ~\text{and} \nonumber\\
&\hspace{4.16cm}x_{a} \cdot x_{b} \geq 0. \label{confim1}
\end{align}

\underline{\emph{Figure \ref{exep}-(ii-5): Small Distance \& Crossing Domains or La}}- \underline{\emph{rge Distances \& Same Domain}:} We note that \{$x_{b} = -0.1$, ${{\underline{x}}_g}  = 0.1$, $x_{a} = 0.8$\} means 1) although George and Alex are in the same domain of opposing the claim but their opposing degrees have large distance, i.e., Alex is much more stubborn while George is much more hesitating, 2) Bob is in the other domain, but he likes George, is very hesitating. In this scenario, George should favor Bob's opinion more.  This expected behavior is described by
\begin{align}
&\underline{c}({{\underline{\mathrm{x}}}_g},x_{b}) > \underline{c}({{\underline{\mathrm{x}}}_g},x_{a}), \hspace{0.12cm}\!~\text{if}~\!| {{{\underline{\mathrm{x}}}_g} \!-\! x_{b}} | = \zeta({\underline{\mathrm{x}}}_g,x_{a},x_{b}) | {{\underline{\mathrm{x}}}_g} \!-\! x_{a}|,  \nonumber\\
&\hspace{1.16cm}\zeta({\underline{\mathrm{x}}}_g,x_{a},x_{b}) < 1, {{\underline{\mathrm{x}}}_g} \cdot x_{b} < 0 ~\text{and}~{{\underline{\mathrm{x}}}_g} \cdot x_{a} > 0. \label{confim0}
\end{align}

\begin{rem}
Taking $\underline{\mathrm{x}}_{g} > 0$ as an example and considering $0 < \zeta({\underline{\mathrm{x}}}_g,x_{a},x_{b}) < 1$, the condition Eq. \eqref{confim0} implies that $\underline{c}({{\underline{\mathrm{x}}}_g},x_{b}) > \underline{c}({{\underline{\mathrm{x}}}_g},x_{a})$. This means that George puts larger influence weight on $x_{b}$ than $x_{a}$ when the ratio of their opinion differences is larger than the threshold, i.e, $\frac{{\left| {\underline{\mathrm{x}}_{g} - x_{a}} \right|}}{{\left| {\underline{\mathrm{x}}_{g} - x_{b}} \right|}} = \frac{1}{\zeta({\underline{\mathrm{x}}}_g,x_{a},x_{b})} > 1$.
\end{rem}

\vspace{-0.10cm}
\subsection{Definition: Asymmetric Cognitive Bias }
Based on the practical asymmetry behavior described in the Section III-A, we first present the formal definition of capturing asymmetric confirmation bias, and then extend the definition to the asymmetric novelty bias.
\begin{defa}
The influence weight $\underline{c}( {{{\underline{\mathrm{x}}}_i}(x,k,\tau_{i}),{x_j}(k)})$ in Eq. \eqref{sdw2b} is said to capture the asymmetric confirmation bias if it satisfies Eq. \eqref{confim4}--\eqref{confim0} simultaneously. \label{ascb}
\end{defa}

Lamberson and Soroka in \cite{lamberson2018model} revealed that negative information -- compared with positive information -- is more ``outlying'', since it is far away from expectations. Inspired by the discovery, an individual's sensed novelty/outlying/surprising degree of information is measured in terms of sensed expectation from her surroundings in memory. For example, if George holds novelty bias, he will prefer news/opinions that has larger distance with her own surrounding expectation ${{\overline{\mathrm{x}}}_g}(x,k,\tau_{g})$. According to the same logic that describes the expected behavior Eq. \eqref{confim4}-\eqref{confim0} due to asymmetric confirmation bias, the expected asymmetry behaviors due to asymmetric novelty bias are formally described by
\begin{align}
&\overline{c}({{\overline{\mathrm{x}}}_g},x_{a}) = \overline{c}({{\overline{\mathrm{x}}}_g},x_{b}), \hspace{0.14cm}~\text{if}~x_{a} = -{{x}_{b}} ~\text{and}~ {{\underline{\mathrm{x}}}_g}  = 0,\label{nconfim4} \\
&\overline{c}({{\overline{\mathrm{x}}}_g},x_{a}) < \overline{c}({{\overline{\mathrm{x}}}_g},x_{b}), \hspace{0.14cm}~\text{if}~| {{{\overline{\mathrm{x}}}_g} - x_{a}} | = | {{\overline{\mathrm{x}}}_g} - x_{b}| ~\text{and} \nonumber\\
&\hspace{3.95cm}
{{\overline{\mathrm{x}}}_g} \cdot x_{a} > 0 \geq {{\overline{\mathrm{x}}}_g} \cdot x_{b}, \label{nconfim3}\\
&\overline{c}({{\overline{\mathrm{x}}}_g},x_{a}) < \overline{c}({{\overline{\mathrm{x}}}_g},x_{b}), \hspace{0.14cm}~\text{if}~| {{{\overline{\mathrm{x}}}_g} - x_{a}} | = | {{\overline{\mathrm{x}}}_g} - x_{b}| ~\text{and} \nonumber\\
&\hspace{3.95cm}
{{\overline{\mathrm{x}}}_g} \cdot x_{a} > {{\overline{\mathrm{x}}}_g} \cdot x_{b} \geq 0, \label{nconfim2}\\
&\overline{c}({{\overline{\mathrm{x}}}_g},x_{b}) < \overline{c}({{\overline{\mathrm{x}}}_g},x_{a}), \hspace{0.14cm}~\text{if}~| {{{\overline{\mathrm{x}}}_g} - x_{b}} | < | {{\overline{\mathrm{x}}}_g} - x_{a}| ~\text{and} \nonumber\\
&\hspace{4.16cm}x_{a} \cdot x_{b} \geq 0, \label{nconfim1}\\
&\overline{c}({{\overline{\mathrm{x}}}_g},x_{b}) < \overline{c}({{\overline{\mathrm{x}}}_g},x_{a}), \hspace{0.12cm}\!~\text{if}~\!| {{{\overline{\mathrm{x}}}_g} \!-\! x_{b}} | = \breve{\zeta}({\overline{\mathrm{x}}}_g,x_{a},x_{b}) | {{\overline{\mathrm{x}}}_g} \!-\! x_{a}|,  \nonumber\\
&\hspace{1.185cm}\breve{\zeta}({\overline{\mathrm{x}}}_g,x_{a},x_{b}) < 1, {{\overline{\mathrm{x}}}_g} \cdot x_{b} < 0 ~\text{and}~{{\overline{\mathrm{x}}}_g} \cdot x_{a} > 0. \label{nconfim0}
\end{align}
The definition of capturing asymmetric novelty bias is then formally presented as follows.
\begin{defa}
The influence weight $\overline{c}( {{{\overline{\mathrm{x}}}_i}(x,k,\tau_{i}),{x_j}(k)})$ in Eq. \eqref{sdw2b} is said to capture the asymmetric novelty bias if it satisfies Eq. \eqref{nconfim4}--\eqref{nconfim0} simultaneously. \label{ascb}
\end{defa}

\subsection{Relevant Work: Confirmation Bias Models}
We now examine the existing models in capturing asymmetric confirmation bias.

\subsubsection{Hegselmann-Krause (HK) Model} The seminal HK model \cite{hegselmann2002opinion}, i.e.,
\begin{subequations}
\begin{align}
  {x_i}\left( {k + 1} \right) &= \frac{1}{{{\left| {{\mathbb{N}_i}\left( k \right)} \right|}}}\sum\limits_{j \in {\mathbb{N}_i}\left( k \right)} {{x_j}\left( k \right)}, ~~~i \in \mathbb{V} \label{conf1gg1} \\
  {\mathbb{N}_i}\left( k \right) &= \left\{ {\left. {j \in \mathbb{V}} \right|{\underline{\varepsilon}_i} \leq \left| {{x_i}\left( k \right) - {x_j}\left( k \right)} \right| \leq {\overline{\varepsilon}_i}} \right\} \label{conf1gg2}
\end{align}\label{conf1gg}
\end{subequations}
\!\!\!has been well recognized for capturing confirmation bias to some extent \cite{del2017modeling,xu2020paradox,del2016spreading}. Depending on the upper confidence level ${\overline{\varepsilon}_i}$ and lower confidence level ${\underline{\varepsilon}_i}$ in Eq. \eqref{conf1gg2}, the HK model \eqref{conf1gg} can partially capture the asymmetric confirmation bias. Concretely, if ${\overline{\varepsilon}_i}$ = ${\underline{\varepsilon}_i}$, the model \eqref{conf1gg} can only capture the symmetric confirmation bias. For example, given ${\overline{\varepsilon}_i}$ = ${\underline{\varepsilon}_i}$ = 0.4 and $x_{i} = 0.2$, if $x_{a} = 0.7$ and $x_{b} = -0.3$, according to the logic \eqref{conf1gg2}, individual $\mathrm{v}_{i}$ will abandon both the opinions $x_{a} = 0.7$ and $x_{b} = -0.3$ completely; if $x_{a} = 0.45$ and $x_{b} = -0.05$, according to Eq. \eqref{conf1gg2}, individual $\mathrm{v}_{i}$ will prefer the opinions $x_{a} = 0.45$ and $x_{b} = -0.05$ equally. Obviously, the model in the scenario of ${\overline{\varepsilon}_i}$ = ${\underline{\varepsilon}_i}$ describes the symmetric behavior.  If ${\overline{\varepsilon}_i} \neq {\underline{\varepsilon}_i}$, the model \eqref{conf1gg} can only partially capture the asymmetric behavior. For example, given ${\overline{\varepsilon}_i}$ = 0.6, ${\underline{\varepsilon}_i}$ = 0.4 and $x_{i} = 0.2$,
\begin{itemize}
  \item Symmetric Behavior: if $x_{a} = 0.6$ and $x_{b} = -0.2$, according to Eq. \eqref{conf1gg2}, individual $\mathrm{v}_{i}$ will also prefer the opinions $x_{a} = 0.6$ and $x_{b} = -0.2$ equally.
  \item Asymmetric Behavior: if $x_{a} = 0.7$ and $x_{b} = -0.2$, individual $\mathrm{v}_{i}$ will completely abandon $x_{a} = 0.7$ and take $x_{b} = -0.2$ into her opinion evolution.
\end{itemize}

\subsubsection{Continuous State-Dependent Influence} To investigate polarization and homogeneity, the DeGroot model with continuous state-dependent influence is proposed in \cite{jabin2014clustering,motsch2014heterophilious}:
\begin{subequations}
\begin{align}
   {x_i}( {k + 1}) &= \frac{\sum\limits_{j \in \mathbb{V}} {c( {{x_i}(k),{x_j}(k)}){x_j}(k)}}{{\sum\limits_{j \in \mathbb{V}} {c( {{x_i}( k ),{x_j}( k)})} }}, ~~i \in \mathbb{V} \label{conf2gg1}\\
  c({x_i},{x_j}) &= \varphi( {{{\left| {{x_i} - x_{j}} \right|}^2}}), ~~~\varphi :\left[ {0,\infty } \right) \to \left[ {0,\infty } \right). \label{conf2gg2}
\end{align}\label{conf2gg}
\end{subequations}
\!\!\!The model can only capture symmetric confirmation bias. For example, given $x_{a} = -0.3$, $x_{b} = 0.5$ and $x_{j} = 0.1$,  according to Eq. \eqref{conf2gg2}, individual $\mathrm{v}_{i}$ exhibits symmetric behavior: $c({x_i},{x_a}) = c({x_i},{x_b})$, since $| {{x_i} - {x_a}} |^{2} = | {{x_j} - {x_b}} |^{2}  = 0.16$.

\vspace{-0.00cm}
\subsection{Asymmetric Cognitive Bias Modeling}
We present the systemic guidance of modeling asymmetric confirmation and novelty bias. In this paper, we construct $\underline{c}( {{{\underline{\mathrm{x}}}_g},{x_a}})$ and $\overline{c}( {{{\overline{\mathrm{x}}}_g},{x_a}})$ to have the following general forms:
\begin{align}
\underline{c}( {{{\underline{\mathrm{x}}}_g},{x_a}}) &= {\underline{\mathfrak{g}}_g}( {{\underline{\mathfrak{f}}_g}(\underline{\mathrm{x}}_{g}) - {\underline{\mathfrak{f}}_g}({x_a})}) \geq 0,\label{conf2}\\
\overline{c}( {{{\overline{\mathrm{x}}}_g},{x_a}}) &= {\overline{\mathfrak{g}}_g}( {{\overline{\mathfrak{f}}_g}({\overline{\mathrm{x}}_g}) - {\overline{\mathfrak{f}}_g}({x_a})}) \geq 0.\label{conf3}
\end{align}

We first present the sufficient and necessary conditions of the model \eqref{conf2} to satisfy \eqref{confim4}-\eqref{confim0} in the following theorem, whose proof appears in \underline{Appendix A}.
\begin{thm}
The influence weight $\underline{c}( {{{\underline{\mathrm{x}}}_g},{x_a}})$ given in \eqref{conf2} satisfies \eqref{confim4}-\eqref{confim0} if and only if
\vspace{-0.10cm}
\begin{align}
&{\underline{\mathfrak{g}}_g}( {{\underline{\mathfrak{f}}_g}({\underline{\mathrm{x}}}_g) - {\underline{\mathfrak{f}}_g}(x_a)})~\text{is strictly decreasing w.r.t. the distance}\nonumber\\
&\hspace{5.20cm}|{{\underline{\mathfrak{f}}_g}({\underline{\mathrm{x}}}_g) - {\underline{\mathfrak{f}}_g}( {{x_a}})} |, \label{confcc1}\\
&{\underline{\mathfrak{f}}_g}(x_a)~\text{is strictly increasing w.r.t. $x_a$}, \label{confcc1ab}\\
&{\underline{\mathfrak{f}}_g}({\underline{\mathrm{x}}}_g) > \frac{{{\underline{\mathfrak{f}}_g}( x_a) + {\underline{\mathfrak{f}}_g}(x_b)}}{2}, ~\text{if}~| {x_a \!-\! {\underline{\mathrm{x}}}_g} | = | {x_b \!-\! {\underline{\mathrm{x}}}_g}|, {x_a} > x_b    \nonumber\\
&\hspace{6.10cm}\text{and}~{\underline{\mathrm{x}}}_g > 0, \label{confcc1ac}\\
&{\underline{\mathfrak{f}}_g}({\underline{\mathrm{x}}}_g) < \frac{{{\underline{\mathfrak{f}}_g}( {x_a}) + {\underline{\mathfrak{f}}_g}(x_b)}}{2}, ~\text{if}~| {x_a \!-\! {\underline{x}}_g} | = | {x_b \!-\! {\underline{\mathrm{x}}}_g}|, x_a < x_b    \nonumber\\
&\hspace{6.10cm}\text{and}~{\underline{\mathrm{x}}}_g < 0, \label{confcc1ad}\\
&\underline{{\mathfrak{f}}}_g(0) = \frac{{{\underline{\mathfrak{f}}_g}(x_a) + {\underline{\mathfrak{f}}_g}({ - x_a})}}{2}. \label{confcc1ae}
\end{align} \label{th1}
\end{thm}

\vspace{-0.65cm}
To end this section, we present the sufficient and necessary conditions of the model \eqref{conf3}  satisfying Eq. \eqref{nconfim4}-\eqref{nconfim0} in the following theorem, whose proof completely follows the proof path of Theorem \ref{th1}, and it is thus omitted.

\begin{thm}
The influence weight $\overline{c}( {{{\overline{\mathrm{x}}}_g},{x_a}})$ given in Eq. \eqref{conf3} satisfies Eq. \eqref{nconfim4}-\eqref{nconfim0} if and only if
\vspace{-0.10cm}
\begin{align}
&{\overline{\mathfrak{g}}_g}( {{\overline{\mathfrak{f}}_g}({\overline{\mathrm{x}}}_g) - {\overline{\mathfrak{f}}_g}(x_a)})~\text{is strictly increasing w.r.t. the distance}\nonumber\\
&\hspace{5.20cm}|{{\overline{\mathfrak{f}}_g}({\overline{\mathrm{x}}}_g) - {\overline{\mathfrak{f}}_g}( {{x_a}})} |, \label{nconfcc1}\\
&{\overline{\mathfrak{f}}_g}(x_a)~\text{is strictly increasing w.r.t. $x_a$}, \label{nconfcc1ab}\\
&{\overline{\mathfrak{f}}_g}({\overline{\mathrm{x}}}_g) > \frac{{{\overline{\mathfrak{f}}_g}( x_a) + {\overline{\mathfrak{f}}_g}(x_b)}}{2}, ~\text{if}~| {x_a \!-\! {\overline{\mathrm{x}}}_g} | = | {x_b \!-\! {\overline{\mathrm{x}}}_g}|, {x_a} > x_b    \nonumber\\
&\hspace{6.10cm}\text{and}~{\overline{\mathrm{x}}}_g > 0, \label{nconfcc1ac}\\
&{\overline{\mathfrak{f}}_g}({\overline{\mathrm{x}}}_g) < \frac{{{\overline{\mathfrak{f}}_g}( {x_a}) + {\overline{\mathfrak{f}}_g}(x_b)}}{2}, ~\text{if}~| {x_a \!-\! {\overline{\mathrm{x}}}_g} | = | {x_b \!-\! {\overline{\mathrm{x}}}_g}|, x_a < x_b    \nonumber\\
&\hspace{6.10cm}\text{and}~{\overline{\mathrm{x}}}_g < 0, \label{nconfcc1ad}\\
&\overline{{\mathfrak{f}}}_g(0) = \frac{{{\overline{\mathfrak{f}}_g}(x_a) + {\overline{\mathfrak{f}}_g}({ - x_a})}}{2}. \label{nconfcc1ae}
\end{align}\label{thk2}
\end{thm}

\vspace{-0.5cm}
\section{Problem \ref{pr2}: M$^{3}$IRL Algorithm}
\vspace{-0.1cm}
To address the Problem \ref{pr2}, we propose the M$^{3}$IRL: a Memorized Model and Maximum-entropy based Inverse Reinforcement Learning with local optimality. M$^{3}$IRL is based on the social information-diffusion model \eqref{kka} and its procedure is described by Algorithm \ref{alg1-16}.

\begin{algorithm} \small{ 
  \caption{M$^{3}$IRL Algorithm} \label{ALG1}
  \KwIn{Trajectory sample \eqref{tras}, basis function library: Eq. \eqref{bc1}--\eqref{bc4}.}
   Fit the social information-diffusion model \eqref{kka} using the trajectory sample \eqref{tras};\\
   Compute the gradient and Hessian according to Eq. \eqref{clm1} and Eq. \eqref{clm2}, respectively;\\
Compute the approximate of conditional probability distribution according to Eq. \eqref{ahhhvg};\\
Compute the preference parameter vectors $\alpha$ and $\theta$ according to Eq. \eqref{mko2} using the library of basis functions Eq. \eqref{bc1}--\eqref{bc4}.
\label{alg1-16}}
\end{algorithm}

\vspace{-0.7cm}
\subsection{Cost Function Formula}
\vspace{-0.1cm}
To formulate the learning problem, we separate the set of individuals $\mathbb{V}$ into the set of humans $\mathbb{H}$ and the set of targets $\mathbb{T}$, i.e., $\mathbb{V} = \mathbb{H} \bigcup \mathbb{T}$ and $\mathbb{H} \bigcap \mathbb{T} = \emptyset$. In light of social information-diffusion model \eqref{kka}, we rewrite the information diffusion dynamics \eqref{kka} in the form:
\begin{align}
\tilde{x}(k + 1) = f(\tilde{x}[(k-\tau):k],~u(k),~s), \label{nfdyna}
\end{align}
where $\tilde{x}[(k-\tau):k] = \left\{ {\tilde{x}(k - \tau ),\tilde{x}(k - \tau  + 1), \ldots ,\tilde{x}(k)} \right\}$, 
\begin{align}
f_i(\tilde{x}[(k\!-\!\tau)\!:\!k],u(k),s) \!=\!  {\alpha _i}(x,k,\tau_{i}){s_i} \!+\! \sum\limits_{j \in \mathbb{V}} \!\!{{c_{ij}}\!(x,k,\tau_{i}){x_j}(k)}, \nonumber
\end{align}
$\tau  = \mathop {\max }\limits_{i \in \mathbb{H}} \left\{ {{\tau _i}} \right\}$, $\tilde{x} \in \mathbb{R}^{|\mathbb{H}|}$ denotes the opinion vector of humans, and $u(k) \in \mathbb{R}^{|\mathbb{T}|}$ denotes the opinion/action vector of targets. For convenience, we let $\mathbb{H} = \{1,2, \ldots ,\left| \mathbb{H} \right|\}$ and $\mathbb{T} = \{\left| \mathbb{H} \right| + 1,\left| \mathbb{H} \right| + 2, \ldots ,\left| \mathbb{H} \right| +\left| \mathbb{T} \right|\}$, such that
\begin{align}
\tilde x\left( k \right) &= {\left[ {{x_1}\left( k \right),{x_2}\left( k \right), \ldots ,{x_{\left| \mathbb{H} \right|}}\left( k \right)} \right]^\top}, \nonumber \\
 u\left( k \right) &= {\left[ {{x_{\left| \mathbb{H} \right| + 1}}\left( k \right),{x_{\left| \mathbb{H} \right| + 2}}\left( k \right), \ldots ,{x_{\left| \mathbb{H} \right| + \left| \mathbb{T} \right|}}\left( k \right)} \right]^\top}.
\end{align}

We observe the public evolving opinions to learn the cost functions of targets.  We denote an observed trajectory in a finite time interval $\{k, k+1, \ldots, k+\mathfrak{l}-1\}$  as
\begin{align}
&\mathfrak{T}(\mathbf{u}) \triangleq \{(\tilde{x}(k+1),u(k)), (\tilde{x}(k+2),u(k+1)), ~\ldots~, \nonumber\\
&\hspace{3.95cm}(\tilde{x}(k + \mathfrak{l}),u(k +\mathfrak{l}-1))\}, \label{tras}
\end{align}
and we also define
\begin{align}
\mathbf{u} &\triangleq \left[ {u(k); ~u( {k + 1}); ~\ldots~; ~u( {k + \mathfrak{l} - 1})} \right] \in \mathbb{R}^{\mathfrak{l}|\mathbb{T}|}, \label{kcm1}\\
\mathbf{x} &\triangleq \left[ {\tilde{x}(k+1); ~\tilde{x}( {k + 2}); ~\ldots~; ~\tilde{x}( {k + \mathfrak{l}})} \right] \in \mathbb{R}^{\mathfrak{l}|\mathbb{H}|}. \label{kcm1ab}
\end{align}

\begin{rem}
The trajectory sample \eqref{tras}  denoted by $\mathfrak{T}(\mathbf{u})$ rather than $\mathfrak{T}(\mathbf{x},\mathbf{u})$ is due to the implication that the evolving opinions $\mathbf{x}$ depend on the actions of observed targets.
\end{rem}

The action space of observed targets is defined as
\begin{align}
 \mathbb{U}^{|\mathbb{T}|} \triangleq [-1,1]^{|\mathbb{T}|} \!\triangleq [-1,1] \times [-1,1] \times \ldots \times [-1,1],\label{kc1}
\end{align}
such that $ u(k) \in  \mathbb{U}^{|\mathbb{T}|}$.

In the framework of cost function learning, we assume each target's cost function denoted by ${r}_{i}(\mathbf{x},\mathbf{u})$  consists of $p$ basis functions:
\begin{subequations}
\begin{align}
{r}_{i}(\mathbf{x}, \mathbf{u}) &= \sum\limits_{q = 1}^p {{\theta _{iq}}} {c_q}( {\mathbf{x},\mathbf{u}}), ~~~~~~~~\text{with}~ \sum\limits_{q = 1}^p |{{\theta _{iq}}}| = 1 \label{pcp}\\
{c_q}( {\mathbf{x},\mathbf{u}}) &\triangleq \sum\limits_{j = k}^{k + \mathfrak{l} - 1}{\breve{c}_q}(\tilde{x}(j),u(j)), ~~q = 1,2, \ldots, p. \label{kcm1abc}
\end{align} \label{mmllnn}
\end{subequations}

\vspace{-0.3cm}
The targets usually do not have identical importance on the public opinion evolution, which is due to, for example, different numbers of their followers. Motivated by this, we impose importance weights on each target's cost function, which can be leveraged to transform multiple cost functions into a single one, i.e.,
\begin{align}
r(\mathbf{u}) \!=\! \sum\limits_{i \in  \mathbb{T}} {{\alpha _i}{r_i}(\mathbf{x},\mathbf{u})}, \text{with}~{\alpha _i} \!\ge\! 0 ~\text{and}~\sum\limits_{i \in  \mathbb{T}} {{\alpha _i} \!=\! 1},\label{eq:dm1}
\end{align}
considering which and \eqref{mmllnn},  we define a set of parameter vectors:
\begin{align}
\alpha &\triangleq [\alpha_{1}, ~\alpha_{2}, ~\ldots, ~\alpha_{{|\mathbb{T}|}-1}, ~\alpha_{{|\mathbb{T}|}}]^{\top} \in \mathbb{R}^{{|\mathbb{T}|}}, \label{ved1}\\
\theta_{i} &\triangleq [\theta_{i1}, ~\theta_{i2}, ~\ldots, ~\theta_{i(p-1)}, ~\theta_{ip}]^{\top}\in \mathbb{R}^{p}, \label{ved2}\\
\theta &\triangleq [\theta_{i}; ~\theta_{2}; ~\ldots; ~\theta_{{|\mathbb{T}|}-1}; ~\theta_{{|\mathbb{T}|}}] \in \mathbb{R}^{{|\mathbb{T}|}p}. \label{ved3}
\end{align}
Hereto, the Problem \ref{pr2} can be reformulated as: \emph{given the basis functions, ${c_q}( {\mathbf{x},\mathbf{u}})$, $q=1,2,\ldots,p$, and the social information-diffusion model \eqref{kka}, inferr the coefficient vectors $\alpha$ and $\theta$ from a single finite-time trajectory sample \eqref{tras}}.

\vspace{-0.5cm}
\subsection{Likelihood via Maximum Entropy}
We let $p(\mathbf{u}| \theta,\alpha,\tilde{x}(k))$ denote the conditional probability distribution of time-series action $\mathbf{u}$, given the coefficient vectors $\alpha$ and $\theta$ and the initial public opinion $\tilde{x}(k)$. The maximum entropy problem is formally formulated as
\vspace{-0.00cm}
\begin{subequations}
\begin{align}
&\mathop {\min }\limits_{p\left( {\mathbf{u}\left| \theta,\alpha,\tilde{x}(k)  \right.} \!\right)} \!\int\limits_{\mathbf{u} \in {\mathbb{U}^{\mathfrak{l}{|\mathbb{T}|}}}} \!\!\!\!\! {p\!\left( {\mathbf{u}\left| \theta,\alpha,\tilde{x}(k)  \right.}\! \right)\ln p\!\left( {\mathbf{u}\left| \theta,\alpha,\tilde{x}(k)  \right.} \!\right)\mathrm{d}\mathbf{u}} \\
\!\!\!\!&\text{subject to}   \nonumber\\
\!\!\!\!&\int\limits_{\mathbf{u} \in {\mathbb{U}^{\mathfrak{l}{|\mathbb{T}|}}}} {p\left( {\mathbf{u}\left| \theta,\alpha,\tilde{x}(k)  \right.} \!\right)\mathrm{d}\mathbf{u}}  = 1,\label{hhh1}\\
\!\!\!\!&\int_{\mathbf{u} \in {\mathbb{U}^{\mathfrak{l}{|\mathbb{T}|}}}} \!\!\!\! {p\!\left( {\left. \mathbf{u} \right|\theta ,\alpha,\tilde{x}(k)} \right)} {c_q}\!\left( {\mathbf{x},\mathbf{u}} \right)\!\mathrm{d}\mathbf{u} \!=\! {\bar c_q}, q \!\in\! \{1,\ldots,p\},
\end{align}\label{men}
\end{subequations}
\!\!\!where ${\bar c_q}$ denotes the expectation of ${c_q}\left( {\mathbf{x},\mathbf{u}} \right)$. The optimal solution is formally presented in the following lemma, whose proof is given in Appendix B in \cite{mao2022cost}.

\begin{lem} \cite{mao2022cost}
The solution of problem \eqref{men} is
\begin{align}
p\!\left( {\mathbf{u} \left| \theta,\alpha,\tilde{x}(k)  \right.} \! \right) &= \frac{{{e^{\sum\limits_{i \in \mathbb{T}} {\sum\limits_{q = 1}^p {{\alpha _i}{\theta _{iq}}{c_q}\left( {\mathbf{x},\mathbf{u}} \right)} } }}}}{{\int_{\tilde{\mathbf{u}} \in {\mathbb{U}^{\mathfrak{l}{|\mathbb{T}|}}}} {{e^{\sum\limits_{i \in \mathbb{T}} {\sum\limits_{q = 1}^p {{\alpha _i}{\theta _{iq}}{c_q}\left( {\tilde{\mathbf{x}},\tilde{\mathbf{u}}} \right)} } }}} \mathrm{d}\tilde{\mathbf{u}}}} \nonumber\\
&= \frac{{{e^{r(\mathbf{u})}}}}{{\int_{\tilde{\mathbf{u}}  \in {\mathbb{U}^{\mathfrak{l}{|\mathbb{T}|}}}} {{e^{{r}(\tilde{\mathbf{u}} )}}} \mathrm{d}\tilde{\mathbf{u}} }}. \label{ahhh1zx}
\end{align}
where \eqref{ahhh1zx} from its previous steps is obtained via considering \eqref{mmllnn} and \eqref{eq:dm1}.\label{kkm}
\end{lem}

\begin{rem}
The maximum-entropy based IRL implies that the objective of collective decision\red{-}making among targets is to maximize the likelihood of the observed sequences of actions and evolving opinions.
\end{rem}

If using only a trajectory of evolving opinions for cost function learning, local optimization of the likelihood of probability distribution is needed, and the algorithm will be model\red{-}based.

For the joint cost function \eqref{eq:dm1}, we, respectively, denote its gradient and Hessian as
\begin{align}
\mathbf{h} &\triangleq \frac{{\partial r\left( \mathbf{u} \right)}}{{\partial \mathbf{u}}}, ~~~~~\mathbf{H} \triangleq \frac{{{\partial ^2}r\left( \mathbf{u} \right)}}{{\partial {\mathbf{u}^2}}}.\label{mmc}
\end{align}
With the defined matrices at hand, we now make the assumptions for deriving main results.
\begin{asm} We assume that
\begin{enumerate}
  \item the derivative $\frac{\partial }{{\partial \mathbf{u}}}\left( {\frac{{\partial \mathbf{x}}}{{\partial \mathbf{u}}}} \right)$ is relatively small such that it can be ignored, i.e., $\frac{\partial }{{\partial \mathbf{u}}}\left( {\frac{{\partial \mathbf{x}}}{{\partial \mathbf{u}}}} \right) \approx \mathrm{O}$,
  \item the integration:
  \begin{align}
\int\limits_{\tilde{\mathbf{u}} \in {\mathbb{U}^{\mathfrak{l}{|\mathbb{T}|}}}}\!\!\!{{e^{{{\tilde{\mathbf{u}}}^ \top }\mathbf{h} - {{\tilde{\mathbf{u}}}^ \top }\mathbf{H}\mathbf{u} + \frac{1}{2}{{\tilde{\mathbf{u}}}^ \top }\mathbf{H}\tilde{\mathbf{u}}}}} \mathrm{d}\tilde{\mathbf{u}} \approx \int\limits_{\tilde{\mathbf{u}} \in {\mathbb{U}^{\mathfrak{l}{|\mathbb{T}|}}}}\!\!\!{{e^{{{\tilde{\mathbf{u}}}^ \top }\mathbf{h} - {{\tilde{\mathbf{u}}}^ \top }\mathbf{H}\mathbf{u} }}} \mathrm{d}\tilde{\mathbf{u}}. \nonumber
\end{align}
 \end{enumerate}
\label{asm1}
\end{asm}

We present the approximation of log\red{-}likelihood of joint action policy \eqref{ahhh1zx} via local optimization in the following theorem, whose proof appears in \underline{Appendix B}.
\begin{thm}
Under the Assumption \ref{asm1}-2), the log likelihood of the conditional probability distribution \eqref{ahhh1zx}, i.e., $\log p\left( {\mathbf{u}\left| \theta  \right.,\alpha, \tilde{x}\left( k \right)} \right)$, is approximated as
\begin{align}
\!\!\!\widetilde{\mathcal{L}}(\theta,\alpha) &=  - \frac{1}{2}{\mathbf{u}^\top}\mathbf{H}\mathbf{u} + {\mathbf{u}^\top}\mathbf{h}   \nonumber\\
&~~~~~~~~~~~~ +   \sum\limits_{i = 1}^{\mathfrak{l}{|\mathbb{T}|}} {\log \left( {\frac{{{{\left[ {\mathbf{h} - \mathbf{H}\mathbf{u}} \right]}_i}}}{{{e^{{{\left[ {\mathbf{h} - \mathbf{H}\mathbf{u}} \right]}_i}}} - {e^{ - {{\left[ {\mathbf{h} - \mathbf{H}\mathbf{u}} \right]}_i}}}}}} \right)}.\label{ahhhvg}
\end{align}\label{tokc}
\end{thm}

\vspace{-0.40cm}
\begin{rem}
Following the same proof path, the approximation of the log\red{-}likelihood of conditional probability distribution was first presented in \cite{2012-cioc} as
\begin{align}
\widetilde{\mathcal{L}}(\theta ,\alpha ) = \frac{1}{2}{\mathbf{u}^\top}\mathbf{H}\mathbf{u} + \frac{1}{2}\log \left| { - \mathbf{H}} \right| - \frac{{\mathfrak{l}{|\mathbb{T}|}}}{2}\log 2\pi, \nonumber
\end{align}
which is derived under an implicit assumption that the decision variables are unbounded, such that the Gaussian integral, i.e., $\int_{ - \infty }^\infty  {{e^{ - {u^2}}}\mathrm{d}u}  = \sqrt \pi$, can be leveraged. However, the approximation via Gaussian integral cannot be applied to the social problem as studied in this paper, since the range of decision variables are constrained to the bounded set $[-1,1]$.
\end{rem}

\vspace{-0.4cm}
\subsection{Computation of $\mathbf{H}$ and $\mathbf{h}$}
Moving forward is the computation of $\mathbf{H}$ and $\mathbf{h}$ for solving the log\red{-}likelihood approximation \eqref{ahhhvg}. Under Assumption \ref{asm1}-1), we obtain from \eqref{eq:dm1} and  \eqref{mmc} that 
\begin{align}
\mathbf{h} &= \sum\limits_{i = 1}^{|\mathbb{T}|} {{\alpha _i}\left( {\frac{{\partial {r_i}(\mathbf{x},\mathbf{u})}}{{\partial \mathbf{u}}} + {{\left( {\frac{{\partial \mathbf{x}}}{{\partial \mathbf{u}}}} \right)^\top} }\frac{{\partial {r_i}(\mathbf{x},\mathbf{u})}}{{\partial \mathbf{x}}}} \right)}, \label{clm1}\\
\mathbf{H} &= \sum\limits_{i = 1}^{\left| \mathbb{T} \right|}\alpha _i {\left(\! {\frac{\partial }{{\partial \mathbf{u}}}\!\left(\! {\frac{{\partial {r_i}\left( {\mathbf{x},\mathbf{u}}\right)}}{{\partial \mathbf{u}}}} \!\right) + {{\left( {\frac{{\partial \mathbf{x}}}{{\partial \mathbf{u}}}} \right)^\top}}\!\frac{\partial }{{\partial \mathbf{u}}}\!\left(\! {\frac{{\partial {r_i}\left({\mathbf{x},\mathbf{u}}\right)}}{{\partial \mathbf{x}}}} \!\right)} \!\right)}\nonumber\\
&= \sum\limits_{i = 1}^{|\mathbb{T}|} {{\alpha _i}} \left( {\frac{{{\partial ^2}{r_i}(\mathbf{u},\mathbf{x})}}{{\partial {\mathbf{u}^2}}}} \right. + {\left( {\frac{{\partial \mathbf{x}}}{{\partial \mathbf{u}}}} \right)^ \top }\frac{{{\partial ^2}{r_i}(\mathbf{u},\mathbf{x})}}{{\partial {\mathbf{x}^2}}}\frac{{\partial \mathbf{x}}}{{\partial \mathbf{u}}} \nonumber\\
&\hspace{1.33cm} \left. { + {{\left( {\frac{{\partial \mathbf{x}}}{{\partial \mathbf{u}}}} \right)^\top}}\frac{{{\partial^2}{r_i}\left( {\mathbf{x},\mathbf{u}} \right)}}{{\partial \mathbf{x} \partial \mathbf{u}}} + \frac{{{\partial ^2}{r_i}\left( {\mathbf{x},\mathbf{u}} \right)}}{{\partial \mathbf{u}\partial \mathbf{x}}}\frac{{\partial \mathbf{x}}}{{\partial \mathbf{u}}}} \right)\!,\label{clm2}
\end{align}
where
\begin{align}
\frac{{\partial \mathbf{x}}}{{\partial \mathbf{u}}} &\triangleq \left[ {\begin{array}{*{20}{c}}
  {\frac{{\partial {\mathbf{x}_1}}}{{\partial {\mathbf{u}_1}}}}&{\frac{{\partial {\mathbf{x}_1}}}{{\partial {\mathbf{u}_2}}}}& \ldots &{\frac{{\partial {\mathbf{x}_1}}}{{\partial {\mathbf{u}_{\mathfrak{l}{|\mathbb{T}|}}}}}} \\
  {\frac{{\partial {\mathbf{x}_2}}}{{\partial {\mathbf{u}_1}}}}&{\frac{{\partial {\mathbf{x}_2}}}{{\partial {\mathbf{u}_2}}}}& \ldots &{\frac{{\partial {\mathbf{x}_2}}}{{\partial {\mathbf{u}_{\mathfrak{l}{|\mathbb{T}|}}}}}} \\
   \vdots & \vdots & \vdots & \vdots  \\
  {\frac{{\partial {\mathbf{x}_{\mathfrak{l}\left| \mathbb{H} \right|}}}}{{\partial {\mathbf{u}_{1}}}}}&{\frac{{\partial {\mathbf{x}_{\mathfrak{l}\left| \mathbb{H} \right|}}}}{{\partial {\mathbf{u}_2}}}}& \ldots &{\frac{{\partial {\mathbf{x}_{\mathfrak{l}\left| \mathbb{H} \right|}}}}{{\partial {\mathbf{u}_{\mathfrak{l}{|\mathbb{T}|}}}}}}
\end{array}} \right]\!\!,\label{kkk} \\
\frac{{{\partial ^2}{r_i}\left( {\mathbf{x},\mathbf{u}} \right)}}{{\partial\mathbf{x} \partial \mathbf{u}}} &\triangleq \left[\!\! {\begin{array}{*{20}{c}}
  {\frac{{{\partial ^2}{r_i}\left( {\mathbf{x},\mathbf{u}} \right)}}{{\partial {\mathbf{x}_1}\partial {\mathbf{u}_1}}}} \!\!&\!\! {\frac{{{\partial ^2}{r_i}\left( {\mathbf{x},\mathbf{u}} \right)}}{{\partial {\mathbf{x}_1}\partial {\mathbf{u}_2}}}} \!\!&\!\! \ldots  \!\!&\!\!  {\frac{{{\partial ^2}{r_i}\left( {\mathbf{x},\mathbf{u}} \right)}}{{\partial {\mathbf{x}_1}\partial {\mathbf{u}_{\mathfrak{l}\left| \mathbb{M} \right|}}}}} \\
  {\frac{{{\partial ^2}{r_i}\left( {\mathbf{x},\mathbf{u}} \right)}}{{\partial {\mathbf{x}_2}\partial {\mathbf{u}_1}}}} \!\!&\!\! {\frac{{{\partial ^2}{r_i}\left( {\mathbf{x},\mathbf{u}} \right)}}{{\partial {\mathbf{x}_2}\partial {\mathbf{u}_2}}}} \!\!&\!\! \ldots \!\!&\!\! {\frac{{{\partial ^2}{r_i}\left( {\mathbf{x},\mathbf{u}} \right)}}{{\partial {\mathbf{x}_2}\partial {\mathbf{u}_{\mathfrak{l}\left| \mathbb{M} \right|}}}}} \\
   \vdots \!\!&\!\! \vdots \!\!&\!\! \vdots \!\!&\!\! \vdots  \\
  {\frac{{{\partial ^2}{r_i}\left( {\mathbf{x},\mathbf{u}} \right)}}{{\partial {\mathbf{x}_{\mathfrak{l}\left| \mathbb{H} \right|}}\partial {\mathbf{u}_1}}}} \!\!&\!\! {\frac{{{\partial ^2}{r_i}\left( {\mathbf{x},\mathbf{u}} \right)}}{{\partial {\mathbf{x}_{\mathfrak{l}\left| \mathbb{H} \right|}}\partial {\mathbf{u}_2}}}} \!\!&\!\! \ldots \!\!&\!\! {\frac{{{\partial ^2}{r_i}\left( {\mathbf{x},\mathbf{u}} \right)}}{{\partial {\mathbf{x}_{\mathfrak{l}\left| \mathbb{H} \right|}}\partial {\mathbf{u}_{\mathfrak{l}\left| \mathbb{M} \right|}}}}}
\end{array}} \!\!\right]\!\!.\label{kkkmmm}
\end{align}

\begin{rem}
Under the Assumption \ref{asm1}-1), the Hessian matrix $\mathbf{H}$ in \cite{2012-cioc} is derived as
\begin{align}
\mathbf{H} = \sum\limits_{i = 1}^{\left| \mathbb{T} \right|} {{\alpha _i}\left( {\frac{{{\partial ^2}{r_i}\left( {\mathbf{x},\mathbf{u}} \right)}}{{\partial {\mathbf{u}^2}}} + {{\left( {\frac{{\partial \mathbf{x}}}{{\partial \mathbf{u}}}} \right)}^\top}\frac{{{\partial ^2}{r_i}\left( {\mathbf{x},\mathbf{u}} \right)}}{{\partial {\mathbf{x}^2}}}\frac{{\partial \mathbf{x}}}{{\partial \mathbf{u}}}} \right)}, \nonumber
\end{align}
which is due to the assumption of Markov Decision Process imposed on system model.
\end{rem}

With the $\mathbf{H}$ and $\mathbf{h}$ at hand, the preference parameter vectors $\alpha$ and $\theta$ can be obtained by solving the following nonlinear constraint optimization problem:
\begin{subequations}
\begin{align}
&\left( {{\theta ^*},{\alpha ^*}} \right) = \mathop {\arg \max }\limits_{\theta ,\alpha } \left\{ \widetilde{\mathcal{L}}(\theta,\alpha)\right\}, \\
& \text{subject to}  \sum\limits_{q = 1}^p {\left| {{\theta _{iq}}} \right| = 1,}~ \sum\limits_{i \in \mathbb{T}} {{\alpha _i} = 1,{\alpha _i} \geqslant 0} ,\forall i \in \mathbb{T}, \
\end{align}\label{mko2}
\end{subequations}
where $\widetilde{\mathcal{L}}(\theta,\alpha)$ is given in \eqref{ahhhvg}.

\vspace{-0.3cm}
\subsection{Basis Functions}
In this section, as an example, we present a library of basis functions that can cover targets' cost functions. We then leverage the proposed M$^{3}$IRL algorithm to infer the associated preference parameter vectors $\theta$ and $\alpha$ of targets' cost functions. The considered basis functions are
\begin{align}
\!\!&{c_1}( \mathbf{x},\mathbf{u} ) = \sum\limits_{t = k}^{k + \mathfrak{l} - 1} \sum\limits^{|\mathbb{H}|}_{j = 1} (1 - \tilde{x}_{j}(t))^2,   \label{bc1}\\
\!\!&{c_2}( \mathbf{x},\mathbf{u} ) = \sum\limits_{t = k}^{k + \mathfrak{l} - 1} \sum\limits^{|\mathbb{H}|}_{j = 1} (1 + \tilde{x}_{j}(t))^2,   \label{bc2}\\
\!\!&{c_3}( \mathbf{x},\mathbf{u} ) = \sum\limits_{t = k}^{k + \mathfrak{l} - 1} \sum\limits^{|\mathbb{H}|}_{j = 1} \tilde{x}^2_{j}(t),   \label{bc3}\\
\!\!&{c_{q - |\mathbb{H}| + 1}}( \mathbf{x},\mathbf{u} ) = \sum\limits_{t = k+1}^{k + \mathfrak{l} - 1} {{{\left( {{u_q}(t) - {u_q}(t-1)} \right)}^2}}, ~~q \!\in\! \mathbb{T}.  \label{bc4}
\end{align}

\vspace{-0.1cm}
\begin{rem}
In minimization, the basis functions \eqref{bc1}--\eqref{bc3} indicate that the objectives of steering the public opinions to $+1$, $-1$ and $0$, respectively. The basis functions \eqref{bc4} imply a behavioral motivation of stubbornness. The implicit motivation representations for decision making can be inferred from the conjunctive $\theta$ and $\alpha$.
\end{rem}

The Eq. \eqref{ahhhvg} and \eqref{mko2} indicate that the inference of preference parameters needs the computations of $\mathbf{h}$ and $\mathbf{H}$. Furthermore, the relations \eqref{clm1} and \eqref{clm2} imply that the computations of $\mathbf{h}$ and $\mathbf{H}$ rely on the computations of  $\frac{{\partial {r_i}(\mathbf{u},\mathbf{x})}}{{\partial \mathbf{x}}}$, $\frac{{\partial {r_i}(\mathbf{u},\mathbf{x})}}{{\partial \mathbf{u}}}$, $\frac{{{\partial ^2}{r_i}(\mathbf{u},\mathbf{x})}}{{\partial {\mathbf{x}^2}}}$, $\frac{{{\partial ^2}{r_i}(\mathbf{u},\mathbf{x})}}{{\partial {\mathbf{u}^2}}}$ and $\frac{{{\partial ^2}{r_i}(\mathbf{u},\mathbf{x})}}{{\partial{\mathbf{x}}, \partial{\mathbf{u}}}}$, which are carried out in Appendix D in \cite{mao2022cost}.

\vspace{-0.0cm}
\section{Empirical Validation}
\begin{figure}
\centering
\includegraphics[scale=0.45]{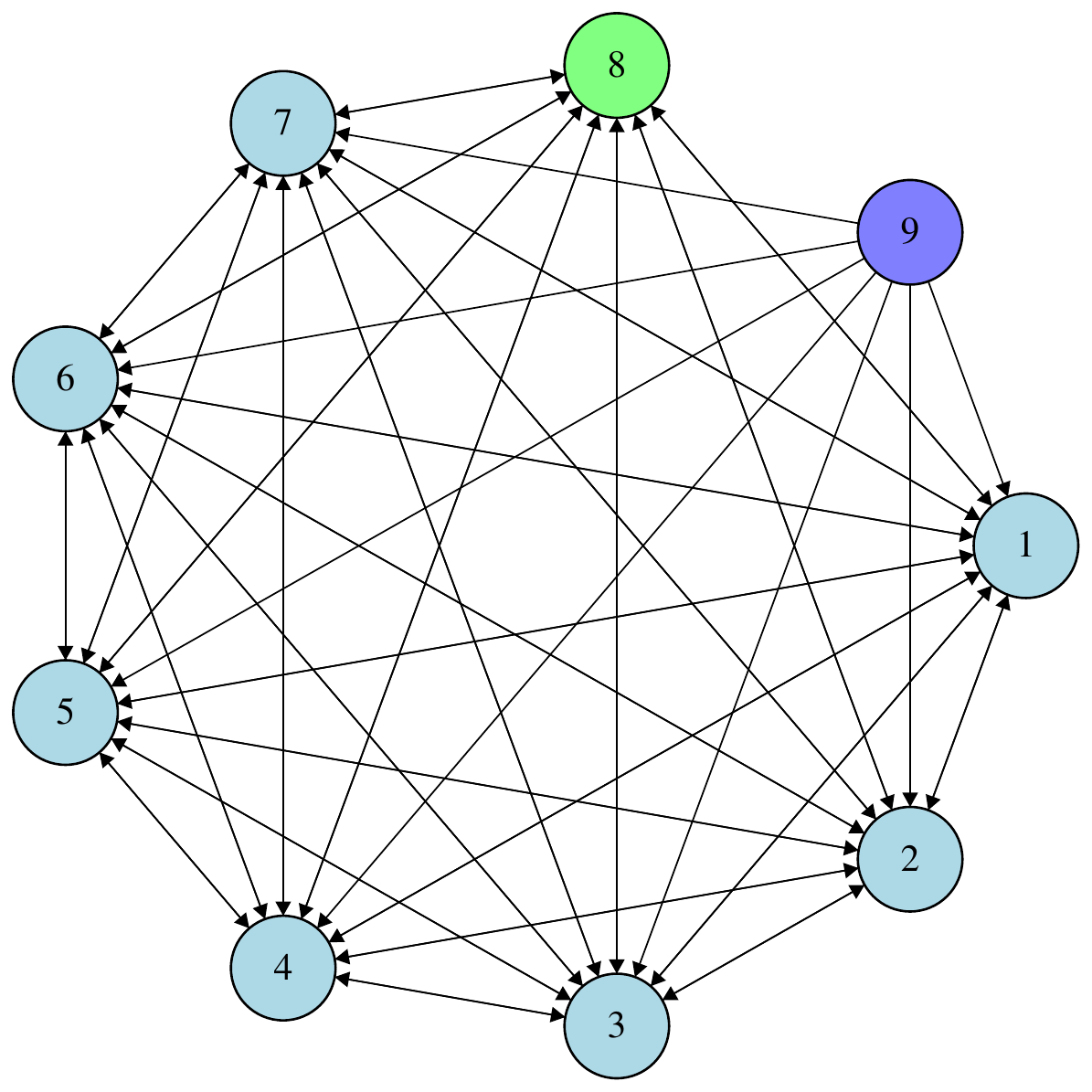}
\caption{Network structure.}
\label{gg}
\end{figure}

We collected nine Twitter users' tweets from January 2021 to September 2021 to validate the effectiveness of the proposed information-diffusion model and the M$^{3}$IRL algorithm. The network structure of nine users is shown in Figure \ref{gg}, where the nodes 8 and 9 are identified as information sources. It follows from  Figure \ref{gg} that $\mathbb{H} = \{1,2,\ldots,7\}$ and $\mathbb{T} = \{8,9\}$.

\vspace{-0.7cm}
\subsection{Data Processing}
The collected tweets are centered around the topic ``COVID-19 Vaccine" and the tweets are sampled biweekly. The tweets are encoded  as numerical values in the range [-1,1], via the BERT Twitter sentiment analysis\footnote{\textcolor[rgb]{1.00,0.00,1.00}{BERT Twitter Sentiment Analysis: https://github.com/OthSay/bert-tweets-analysis}}, which is trained using the Sentiment140 dataset with 1.6 million tweets\footnote{\textcolor[rgb]{1.00,0.00,1.00}{Sentiment140 dataset: https://www.kaggle.com/kazanova/sentiment140}}. The collected tweets with encoded values are available at \textcolor[rgb]{1.00,0.00,1.00}{Twitter$\_$Vaccine$\_$data.xlsx: https://github.com/ymao578/Social-Data}. The encoded $x \in [-1, 0)$ denotes the opinion of opposing COVID-19 Vaccine, $x = 0$ is the neutral opinion, and $x \in (0, 1]$ represents the opinion of supporting COVID-19 Vaccine. A few examples of encoded tweets are given as follows:\\
-- \textit{`Scientists' and `Doctors' at The Lancet and Nature Have Blood on Their Evil Hands} $\Rightarrow$ $-1$.\\
-- \textit{\#COVID19 \#CDCExperts  Is this true? Claim `Falling Ill Or `Death After Receiving COVID Vaccine `Predictable And `Actually A Good Thing
} $\Rightarrow$ $-0.08$.\\
-- \textit{You think that I am against vaccination, I don't. I don't either endorse it nor against it. I don't like the way of handling the data.} $\Rightarrow$ $0$.\\
-- \textit{I am pro (some)vaccines but they need to be tested for 7-10 years.  They need to go through the proper testing. I want long-term data on vaccine outcomes. MRNA technology is synthetic how can that be good for your body?} $\Rightarrow$ $0.216666667$.

\vspace{-0.0cm}
\subsection{Model Fitting}
\vspace{-0.0cm}
The considered decaying-influence model of memory is a simplified version of base level function in the ACT-R declarative memory model \cite{stanley2016comparing,cck}:
\begin{align}
m_{i}(v) =\log (v^{-d_{i}}+1),~~~~v \in \left\{ {1,2, \ldots ,{\tau_i}} \right\},~{d_i} > 0, \label{mim}
\end{align}
where $\tau_i$ is individual $\mathrm{v}_{i}$'s memory horizon and ${d_i}$ is the fitting parameter from real data.

To simplify the model fitting, we ignore humans' innate opinions in model and let $\overline{\mathrm{x}}_{1} = \overline{\mathrm{x}}_{2} = \ldots = \overline{\mathrm{x}}_{7} = \overline{\mathrm{x}}$ and $d_{1} = d_{2} = \ldots = d_{7} = d$. According to  Theorems \ref{th1} and \ref{thk2}, the social influence models that aim to capture the asymmetric confirmation bias and novelty bias are chosen as
\begin{align}
&\underline{c}( {{{\underline{\mathrm{x}}}_i},{x_j}}) \nonumber\\
&\!=\!  \frac{|\tanh(\underline{\mathrm{x}}_{i}) \!-\! \tanh(x_{j})|^{-\alpha_{i}}}{\sum\limits_{j \in \mathbb{V}} \!\!|\tanh(\underline{\mathrm{x}}_{i}) \!-\! \tanh(x_{j})|^{-\underline{\alpha}_{i}} \!+\! \sum\limits_{j \in \mathbb{V}} \!\! |\tanh(\overline{\mathrm{x}}_{i}) \!-\! \tanh(x_{j})|^{\underline{\alpha}_{i}}},  \nonumber\\
&\overline{c}( {{{\overline{\mathrm{x}}}_i},{x_j}})  \nonumber\\
&\!=\!  \frac{|\tanh(\overline{\mathrm{x}}_{i}) \!-\! \tanh(x_{j})|^{-\alpha_{i}}}{\sum\limits_{j \in \mathbb{V}} \!\!|\tanh(\underline{\mathrm{x}}_{i}) \!-\! \tanh(x_{j})|^{\underline{\alpha}_{i}} \!+\! \sum\limits_{j \in \mathbb{V}} \!\! |\tanh(\overline{\mathrm{x}}_{i}) \!-\! \tanh(x_{j})|^{\underline{\alpha}_{i}}},  \nonumber
\end{align}
where $\alpha_{i} > 0$ is the fitting parameter. Under the settings, according to the dynamics \eqref{kka} and the relation \eqref{sdw2b},  the model \eqref{nfdyna}, without consideration of innate opinions, is rewritten as
\begin{align}
{x_i}(k + 1)
&= \sum\limits_{j \in\mathbb{H}} {{c_{ij}}(x,k,{\tau _i}){x_j}(k)}  + \sum\limits_{j \in \mathbb{T}} {{c_{ij}}(x,k,{\tau _i}){u_j}(k)}, \nonumber
\end{align}
where $i \in \mathbb{H}$. To fit the model from real data, the considered loss function is $e = \sum\limits_{k = 1}^{18} {\left\| {x\left( k \right) - \hat x\left( k \right)} \right\|_2^2}$, where $\hat x\left( k \right)$ denotes the real data of opinion  at time $k$ (given $\hat x\left( 1 \right) = x\left( 1 \right)$).


We let $\tau_{i} = 2, \forall i \in \mathbb{H}$, which means that the fitted model assumes all  the humans' memory horizons are 1 month due to the biweekly sampling rate. The fitted model parameters are summarized as
\begin{itemize}
  \item $\overline{\mathrm{x}} = -1$ and $d = 6.01$.
  \item $\underline{\alpha}_{1} = 1.8$, $\underline{\alpha}_{2} = 2.2$, $\underline{\alpha}_{3} = 1.4$, $\underline{\alpha}_{4} = 2.2$, $\underline{\alpha}_{5} = 0.2$, $\underline{\alpha}_{6} = 1$ and $\underline{\alpha}_{7} = 2.2$.
\end{itemize}

\vspace{-0.8cm}
\subsection{Cost Functions Learning via M$^{3}$IRL}
We use the most recent three data (sampling over 1.5 months: mid-August to end-September) to learn the cost functions. With the encoded data and fitted model parameters, we obtain from Appendix D in \cite{mao2022cost} that
\begin{align}
&\mathbf{u}^{\top}\mathbf{h} = 0.3662\left( {{\alpha _1}{\theta _{11}} + {\alpha _2}{\theta _{21}}} \right) - 0.4609\left( {{\alpha _1}{\theta _{12}} + {\alpha _2}{\theta _{22}}} \right) \nonumber\\
&~~~~~~~~~- 0.0473\left( {{\alpha _1}{\theta _{13}} + {\alpha _2}{\theta _{23}}} \right) - 0.4166\left( {{\alpha _1}{\theta _{14}} + {\alpha _2}{\theta _{24}}} \right), \nonumber\\
&\mathbf{u}^{\top}\mathbf{H}\mathbf{u} = 0.027({\alpha _1} {\theta _{11}} + {\alpha _1}{\theta _{12}} + {\alpha _1}{\theta _{13}} + {\alpha _2}{\theta _{21}} + {\alpha _2}{\theta _{22}} \nonumber\\
&~~~~~~~~~~~~ + {\alpha _2}{\theta _{23}}) + 0.7351\left( {{\alpha _1}{\theta _{14}} + {\alpha _2}{\theta _{24}}} \right), \nonumber\\
&\mathcal{G}\left( {\mathbf{h},\mathbf{H}} \right) = \sum\limits_{i = 1}^4 {\log \left( {\frac{{{\varpi _i}}}{{{e^{{\varpi _i}}} - {e^{ - {\varpi _i}}}}}} \right)}, \nonumber
\end{align}
where
\begin{align}
{\varpi _1} = &-{\text{0.8512}}\left( {{\alpha _1}{\theta _{11}} + {\alpha _2}{\theta _{21}}} \right) + {\text{1}}{\text{.7902}}\left( {{\alpha _1}{\theta _{12}} + {\alpha _2}{\theta _{22}}} \right) \nonumber\\
&- {\text{0}}{\text{.0305}}\left( {{\alpha _1}{\theta _{13}} + {\alpha _2}{\theta _{23}}} \right) + {\text{1}}{\text{.3636}}\left( {{\alpha _1}{\theta _{14}} + {\alpha _2}{\theta _{24}}} \right), \nonumber\\
{\varpi _2} = &-{\text{0.5547}}\left( {{\alpha _1}{\theta _{11}} + {\alpha _2}{\theta _{21}}} \right) + {\text{0}}{\text{.9536}}\left( {{\alpha _1}{\theta _{12}} + {\alpha _2}{\theta _{22}}} \right) \nonumber\\
&+ {\text{0}}{\text{.1994}}\left( {{\alpha _1}{\theta _{13}} + {\alpha _2}{\theta _{23}}} \right) - {\text{1}}{\text{.3636}}\left( {{\alpha _1}{\theta _{14}} + {\alpha _2}{\theta _{24}}} \right), \nonumber \\
{\varpi _3} = &-{\text{0.1860}}\left( {{\alpha _1}{\theta _{11}} + {\alpha _2}{\theta _{21}}} \right) + {\text{0}}{\text{.2281}}\left( {{\alpha _1}{\theta _{12}} + {\alpha _2}{\theta _{22}}} \right) \nonumber\\
&+ {\text{0}}{\text{.0211}}\left( {{\alpha _1}{\theta _{13}} + {\alpha _2}{\theta _{23}}} \right) - {\text{1}}{\text{.3636}}\left( {{\alpha _1}{\theta _{14}} + {\alpha _2}{\theta _{24}}} \right), \nonumber\\
{\varpi _4} = &-{\text{0.5622}}\left( {{\alpha _1}{\theta _{11}} + {\alpha _2}{\theta _{21}}} \right) + {\text{0}}{\text{.6720}}\left( {{\alpha _1}{\theta _{12}} + {\alpha _2}{\theta _{22}}} \right) \nonumber\\
&+ {\text{0}}{\text{.0549}}\left( {{\alpha _1}{\theta _{13}} + {\alpha _2}{\theta _{23}}} \right) - {\text{1}}{\text{.3636}}\left( {{\alpha _1}{\theta _{14}} + {\alpha _2}{\theta _{24}}} \right).\nonumber
\end{align}

\begin{figure}
\centering
\includegraphics[scale=0.22]{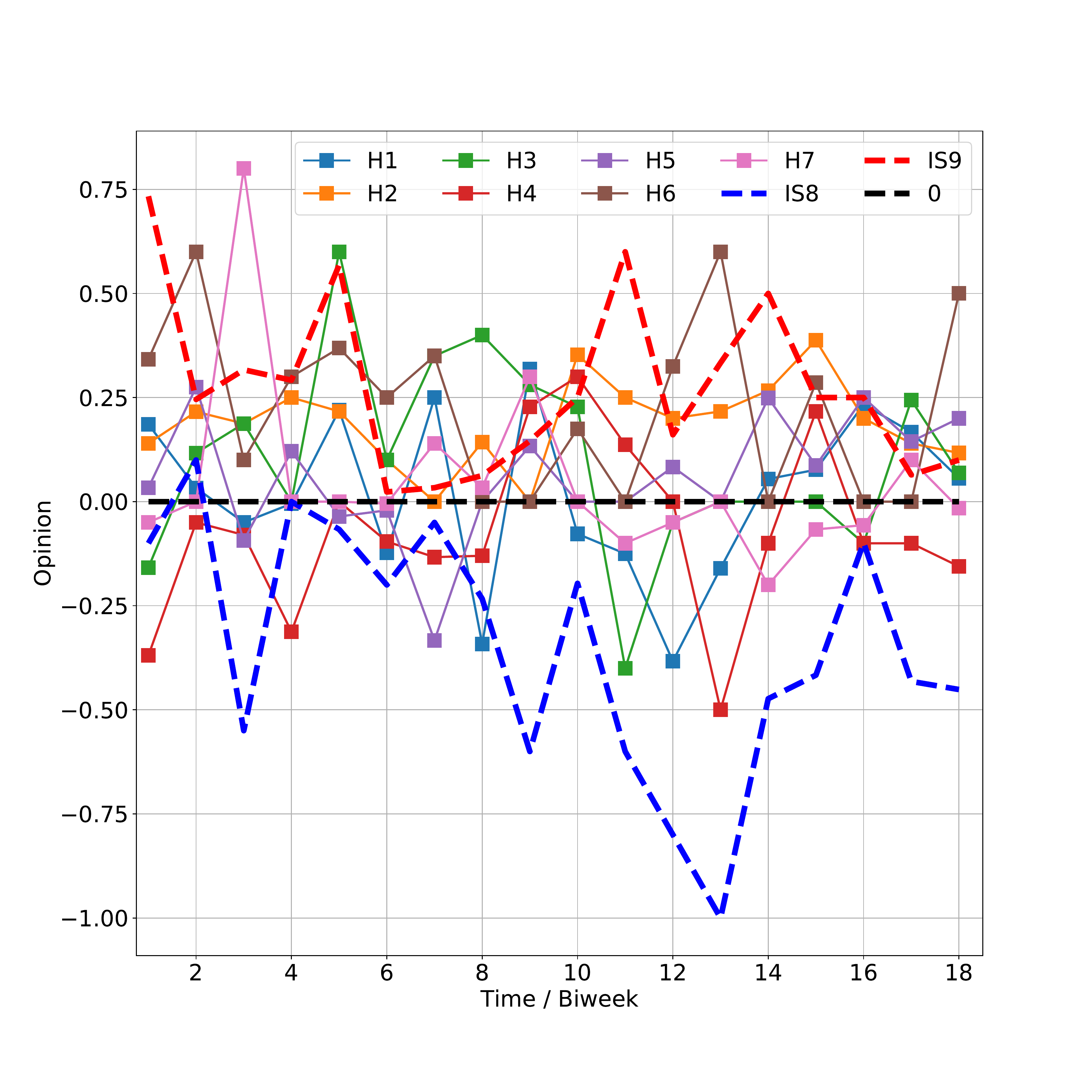}
\caption{Opinion trajectories of ground truth: H1, $\ldots$, H7 denote human 1, $\ldots$, human 7, IS8 and IS9 denote information sources 8 and 9,  respectively.}
\label{real}
\end{figure}

The preference coefficients are obtained through solving Eq. \eqref{mko2} via the constraint optimization toolbox `fmincon' of MATLAB:
\begin{subequations}
\begin{align}
&{\alpha _8} = 0.2040,~~~{\alpha _9} = 0.7960,  \\
&\left[ {{\theta _{81}},{\theta _{82}},{\theta _{83}},{\theta _{84}}} \right] = \left[ { - {\text{0}}{\text{.0852}},0, - {\text{0}}{\text{.1799}},  {\text{0}}{\text{.735}}} \right], \\
&\left[ {{\theta _{91}},{\theta _{92}},{\theta _{93}},{\theta _{94}}} \right] = \left[ {{\text{0}}{\text{.7815}},0,0, - 0.2185} \right]. 
\end{align} \label{ghoooo}
\end{subequations}
With the basis functions \eqref{bc1}--\eqref{bc4}, the learned cost functions of two information sources are
\begin{align}
{r_8}\left( {\mathbf{x},\mathbf{u}} \right) =  &- 0.0852\sum\limits_{t = k}^{k + \mathfrak{l} - 1} {\sum\limits_{j = 1}^{\left| \mathbb{H} \right|} {{{\left( {1 - {x_j}\left( t \right)} \right)}^2} } } \nonumber\\
&+ 0.735 \sum\limits_{t = k}^{k + \mathfrak{l} - 1} {\sum\limits_{j = 1}^{\left| \mathbb{H} \right|} {{{\left( {1 + {x_j}\left( t \right)} \right)}^2}} }  \nonumber\\
& - {\text{0}}{\text{.1799}} \sum\limits_{t = k + 1}^{k + \mathfrak{l} - 1} {{{\left( {{u_8}\left( t \right) - {u_8}\left( t \right)} \right)}^2},}  \label{ck11} \\
{r_9}\left( {\mathbf{x},\mathbf{u}} \right) = &~0.7851\sum\limits_{t = k}^{k + \mathfrak{l} - 1} {\sum\limits_{j = 1}^{\left| \mathbb{H} \right|} {{{\left( {1 - {x_j}\left( t \right)} \right)}^2}} }  \nonumber\\
&- 0.2185\sum\limits_{t = k + 1}^{k + \mathfrak{l} - 1} {{{\left( {{u_9}\left( t \right) - {u_9}\left( t \right)} \right)}^2}} . \label{ck22}
\end{align}

\begin{figure}
\centering
\includegraphics[scale=0.35]{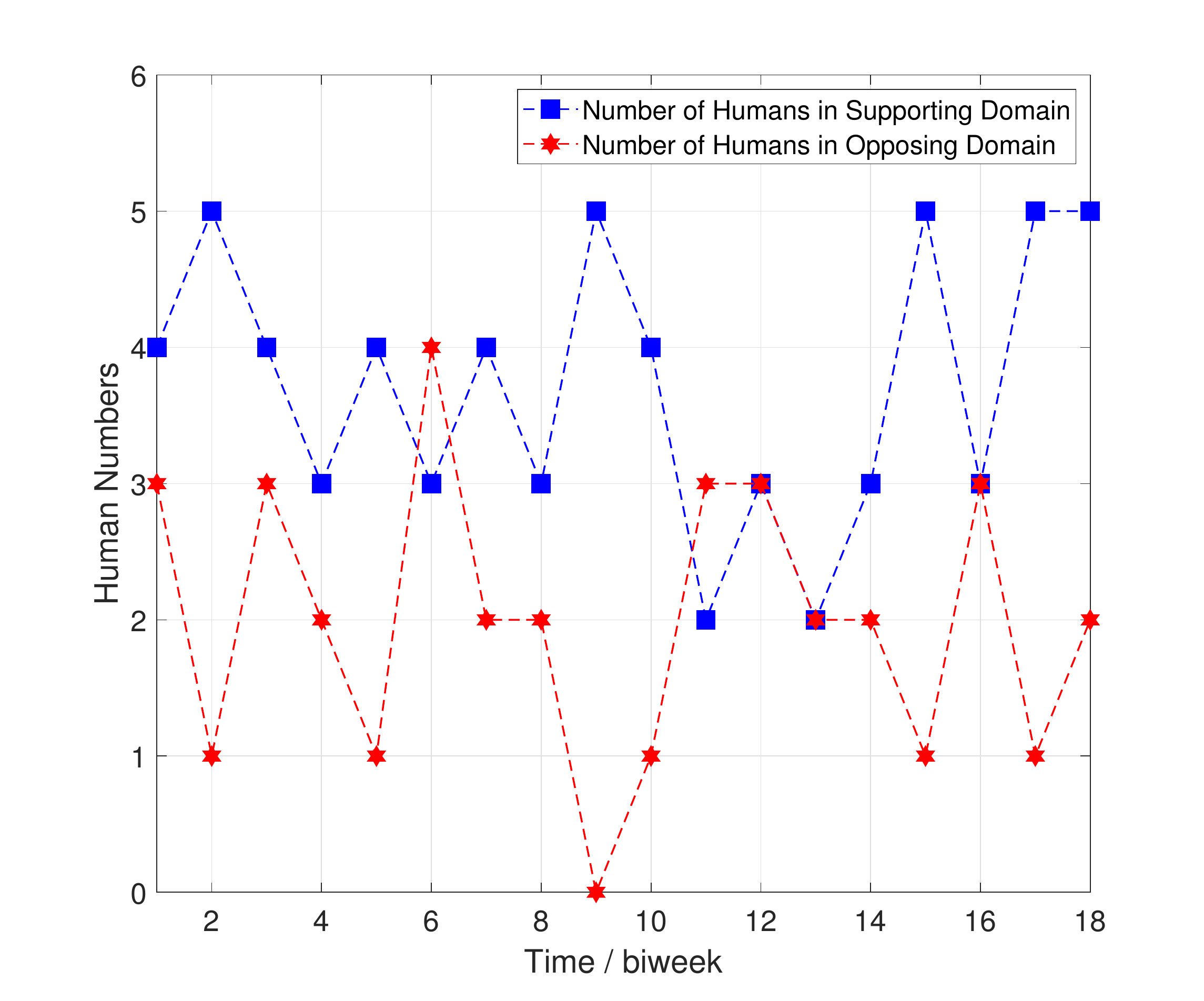}
\caption{Number of humans in supporting domain v.s.  number of humans in the opposing domain. }
\label{realhn}
\end{figure}

Through observing the learned coefficient $\alpha_8$ and $\alpha_9$ and cost functions \eqref{ck11} and \eqref{ck22}, we infer:\\
--  The information source IS8 aims at manipulating the opinions of the public to be against the COVID-19 Vaccine (i.e, $x(k) \in [-1,0)^7$), which is indicated by the first two terms in the right-hand of Eq. \eqref{ck11}. This inference can be demonstrated by IS8's evolving opinions in Figure \ref{real}. \\
--  The information source IS9 aims at leading the opinions of the public to support the COVID-19 Vaccine (i.e, $x(k) \in (0, 1]^7$), which is indicated by the first term in the right-hand of Eq. \eqref{ck22}. This inference can be demonstrated by SI9's evolving opinions in Figure \ref{real}. \\
-- Both information sources IS8 and IS9 prefer to spread outlying opinions or dislike spreading persistent opinions, which is implied by the third and second terms in the right-hands of Eq. \eqref{ck11} and \eqref{ck22}, respectively. The inference can be partially demonstrated by the sharp jumping opinions in the trajectories of IS8 and IS9 in Figure \ref{real}. \\
-- The ${\alpha _9} = 0.7960$ $>$ ${\alpha _8} = 0.2040$ implies that the information source IS9 has a bigger influence than the information source IS8 on the evolving opinions of the observed social community. The inference can be partially demonstrated by the observation of Figure \ref{real} in conjunction with Figure \ref{realhn} that more evolving opinions are in the supporting range $(0,1]$ than the opposing range $[-1,0)$, and IS9 supports COVID-19 vaccine.

\vspace{-0.1cm}
\subsection{Comparision}
\vspace{-0.1cm}
We use the most recent four action data (sampling over 2 months: early-August to end-September) to learn the cost functions. Following the same step to obtain \eqref{ghoooo}, we have
\begin{align}
&{\alpha _8} = 0.2086,~~~{\alpha _9} = 0.7914,  \nonumber\\
&\left[ {{\theta _{81}},{\theta _{82}},{\theta _{83}},{\theta _{84}}} \right] = \left[ { - {\text{0}}{\text{.0806}}, 0.0001, - {\text{0}}{\text{.1853}}, {\text{0}}{\text{.7340}}} \right], \nonumber\\
&\left[ {{\theta _{91}},{\theta _{92}},{\theta _{93}},{\theta _{94}}} \right] = \left[ {{\text{0}}{\text{.7843}},0,0, - 0.2156} \right]. \nonumber
\end{align} 
\vspace{-0.0cm}
With the basis functions \eqref{bc1}--\eqref{bc4}, the learned cost functions of two information sources are
\begin{align}
{r_8}\left( {\mathbf{x},\mathbf{u}} \right) =  &- 0.0806\sum\limits_{t = k}^{k + \mathfrak{l} - 1} {\sum\limits_{j = 1}^{\left| \mathbb{H} \right|} {{{\left( {1 - {x_j}\left( t \right)} \right)}^2} } } \nonumber\\
&+ 0.734 \sum\limits_{t = k}^{k + \mathfrak{l} - 1} {\sum\limits_{j = 1}^{\left| \mathbb{H} \right|} {{{\left( {1 + {x_j}\left( t \right)} \right)}^2}} }  \nonumber\\
&+ 0.0001  \sum\limits_{t = k+1}^{k + \mathfrak{l} - 1} {{{\left( {{u_q}(t) - {u_q}(t-1)} \right)}^2}}  \nonumber\\
& - {\text{0}}{\text{.1853}} \sum\limits_{t = k + 1}^{k + \mathfrak{l} - 1} {{{\left( {{u_8}\left( t \right) - {u_8}\left( t \right)} \right)}^2},}  \nonumber \\
{r_9}\left( {\mathbf{x},\mathbf{u}} \right) = &~0.7843\sum\limits_{t = k}^{k + \mathfrak{l} - 1} {\sum\limits_{j = 1}^{\left| \mathbb{H} \right|} {{{\left( {1 - {x_j}\left( t \right)} \right)}^2}} }  \nonumber\\
&- 0.2156\sum\limits_{t = k + 1}^{k + \mathfrak{l} - 1} {{{\left( {{u_9}\left( t \right) - {u_9}\left( t \right)} \right)}^2}}, \nonumber
\end{align}
which in conjunction with \eqref{ck11} and \eqref{ck22} indicate that the learned cost functions vary with length of finite-time action trajectory. This also means the learned cost functions via our proposed M$^3$ are dynamic, which can capture the influence of dynamic external stimulus on decision making of information sources.

\section{Conclusion}
In this paper, we have proposed the social information-diffusion model which explicitly takes human memory, asymmetric confirmation bias and asymmetric novelty bias into account. Based on the proposed model, we have proposed the M$^{3}$IRL algorithm to learn the cost functions of target individuals. Real data validations suggest the effectiveness of the derived M$^{3}$IRL algorithm and the proposed public opinion evolution model.

As a part of future research, we will investigate the generalization of the cost function learning framework for large-scale social networks with the incorporation of communication detection and classification.

\appendices
\section*{Appendix A: Proof of Theorem \ref{th1}}
\underline{\emph{Sufficient Condition:}} Without loss of generality, we let $x_a > 0$. With the consideration of Eq. \eqref{confcc1ab}, the condition \eqref{confcc1ae} is equivalent to
\begin{align}
|{\underline{\mathfrak{f}}_g}(x_a) - \underline{{\mathfrak{f}}}_g(0)| = {\underline{\mathfrak{f}}_g}(x_a) - \underline{{\mathfrak{f}}}_g(0) &= \underline{{\mathfrak{f}}}_g(0) - {\underline{\mathfrak{f}}_g}(-x_a) \nonumber\\
&= |{\underline{\mathfrak{f}}_g}(-x_a) - \underline{{\mathfrak{f}}}_g(0)|, \nonumber
\end{align}
which, in conjunction with Eq. \eqref{confcc1} and \eqref{confcc1ab}, leads to the \underline{behavior \eqref{confim4}}.

We now consider the condition
\begin{align}
| {{\underline{\mathrm{x}}_g} - {{x}_a}} | = | {\underline{\mathrm{x}}_g - {{x}_b}}|~\text{and}~{x_a}\underline{\mathrm{x}}_g > {x_b}\underline{\mathrm{x}}_g, \label{cmk3as}
\end{align}
which is the union of conditions in Eq. \eqref{confcc1ac} and \eqref{confcc1ad}. If $\underline{\mathrm{x}}_g > 0$, the relationship in Eq. \eqref{cmk3as} implies that ${x_a} - \underline{\mathrm{x}}_g = \underline{\mathrm{x}}_g - {{x}_b} > 0$ and ${x_a} > {x_b}$, which follows from Eq. \eqref{confcc1ab} and \eqref{confcc1ac} that implies $| {{\underline{\mathfrak{f}}_i}( {{x_a}}) - {\underline{\mathfrak{f}}_i}(\underline{\mathrm{x}}_g)} | = {\underline{\mathfrak{f}}_i}( {{x_a}}) - {\underline{\mathfrak{f}}_i}(\underline{\mathrm{x}}_g)$ $<$ $| {{\underline{\mathfrak{f}}_i}(\underline{\mathrm{x}}_g) - {\underline{\mathfrak{f}}_i}({x}_b)}| = {\underline{\mathfrak{f}}_i}(\underline{\mathrm{x}}_g) - {\underline{\mathfrak{f}}_i}({x}_b)$. We then can obtain from Eq. \eqref{confcc1} and \eqref{conf2} that
\begin{align}
&\underline{c}(\underline{\mathrm{x}}_{g},x_{a}) > \underline{c}( {\underline{\mathrm{x}}_{g},{x_{b}}}), ~\text{if}\!~|x_{b} - \underline{\mathrm{x}}_{g}| = |\underline{\mathrm{x}}_{g} - x_{a}|,  {x}_{a} > {x_{b}}, \nonumber\\ &\hspace{6.00cm}\text{and}~\underline{\mathrm{x}}_{g}> 0. \label{cmk3as1}
\end{align}
If $\underline{\mathrm{x}}_g < 0$, with the consideration of Eq. \eqref{confcc1ad}, following the same steps to derive Eq. \eqref{cmk3as1}, we have
\begin{align}
&\underline{c}(\underline{\mathrm{x}}_{g},x_{a}) > \underline{c}( {\underline{\mathrm{x}}_{g},{x_{b}}}), ~\text{if}\!~|x_{b} - \underline{\mathrm{x}}_{g}| = |\underline{\mathrm{x}}_{g} - x_{a}|,  {x}_{a} < {x_{b}}, \nonumber\\ &\hspace{6.00cm}\text{and}~\underline{\mathrm{x}}_{g}< 0. \label{cmk3as2}
\end{align}

We note that the union of Eq. \eqref{cmk3as1} and \eqref{cmk3as2} is equivalent to
\begin{align}
&\underline{c}(\underline{\mathrm{x}}_{g},x_{a}) > \underline{c}( {\underline{\mathrm{x}}_{g},{x_{b}}}), ~\text{if}\!~|x_{b} - \underline{\mathrm{x}}_{g}| = |\underline{\mathrm{x}}_{g} - x_{a}| ~\text{and}, \nonumber\\ &\hspace{5.00cm}{x}_{a} \cdot \underline{\mathrm{x}}_{g} >  {x}_{b} \cdot \underline{\mathrm{x}}_{g}. \label{cmk3asun}
\end{align}
Meanwhile, it is straightforward to observe from Eq. \eqref{confim3} and \eqref{confim2} that the union of them is also equivalent to \eqref{cmk3asun}. We thus conclude that the conjunctive conditions Eq. \eqref{confcc1}-\eqref{confcc1ad} lead to the \underline{behavior \eqref{confim3} and \eqref{confim2}}.

Without loss of generality, we let $\underline{\mathrm{x}}_{g} \geq x_a \geq 0$. It follows from Eq. \eqref{confcc1ab} that $| {{\underline{\mathfrak{f}}_g}(\underline{\mathrm{x}}_{g}) - {\underline{\mathfrak{f}}_i}( {{x_a}})} | = {{\underline{\mathfrak{f}}_g}(\underline{\mathrm{x}}_{g}) - {\underline{\mathfrak{f}}_g}( {{x_a}})}$. If $x_a$ decreases to $x_b \geq 0$, we then have $| {{\underline{\mathfrak{f}}_i}(\underline{\mathrm{x}}_{g}) - {\underline{\mathfrak{f}}_i}( {{x_a}})} | < | {{\underline{\mathfrak{f}}_g}(\underline{\mathrm{x}}_{g}) - {\underline{\mathfrak{f}}_i}( {{x_b}})} |$ and $|\underline{\mathrm{x}}_{g} - x_a| < |\underline{\mathrm{x}}_{g} - x_b|$. Considering Eq. \eqref{confcc1} and \eqref{conf2}, we have $\underline{c}(\underline{\mathrm{x}}_{g},x_a) > \underline{c}( {\underline{\mathrm{x}}_{g},x_b})$. If $x_a$ increases to $x_b$ such that $x_b - \underline{\mathrm{x}}_{g}$ $>$ $\underline{\mathrm{x}}_{g} - x_a \geq 0$, we obtain from Eq. \eqref{confcc1ab} that $| {{\underline{\mathfrak{f}}_i}(\underline{\mathrm{x}}_{g}) - {\underline{\mathfrak{f}}_g}( {{x_a}})} | < | {{\underline{\mathfrak{f}}_g}(\underline{\mathrm{x}}_{g}) - {\underline{\mathfrak{f}}_g}( {{x_b}})} |$ and $|\underline{\mathrm{x}}_{g} - x_{a}| < |\underline{x}_{g} - x_{b}|$. Considering Eq. \eqref{confcc1} and \eqref{conf2}, we then have $\underline{c}(\underline{\mathrm{x}}_{g},x_{a}) > \underline{c}( {\underline{\mathrm{x}}_{g},x_b})$. We thus conclude that the conjunctive conditions Eq. \eqref{confcc1} and \eqref{confcc1ab} imply
\begin{align}
&\underline{c}(\underline{\mathrm{x}}_{g},x_{a}) > \underline{c}( {\underline{\mathrm{x}}_{g},\!{x_{b}}}), ~\text{if}~|x_{a} - \underline{\mathrm{x}}_{g}| < |\underline{\mathrm{x}}_{g} - x_{b}|, ~x_{a} \geq 0, \nonumber\\
&\hspace{6.00cm}\text{and}~x_{b} \geq 0. \label{ppk1}
\end{align}
In the case of $0 > \underline{\mathrm{x}}_{g} \geq x_{a}$, following the same steps to derive Eq. \eqref{ppk1}, we have
\begin{align}
&\underline{c}(\underline{\mathrm{x}}_{g},x_{a}) > \underline{c}( {\underline{\mathrm{x}}_{g},\!{x_{b}}}), ~\text{if}~|x_{a} - \underline{\mathrm{x}}_{g}| < |\underline{\mathrm{x}}_{g} - x_{b}|, ~x_{a} \leq 0, \nonumber\\
&\hspace{6.00cm}\text{and}~x_{b} \leq 0. \label{ppk2}
\end{align}
The results \eqref{ppk1} and \eqref{ppk2} indicate that the conjunctive conditions \eqref{confcc1} and \eqref{confcc1ab} result in the \underline{behavior \eqref{confim1}}.

Let us consider the condition
\begin{align}
\underline{\mathrm{x}}_{g} \cdot \breve{x}_b < 0 ~\text{and}~\underline{\mathrm{x}}_{g} \cdot x_a > 0. \label{cmk3as3}
\end{align}
If $\underline{\mathrm{x}}_g > 0$, the condition \eqref{cmk3as3} implies that $\breve{x}_b < 0$ and $x_a > 0$. Without loss of generality, we let $\underline{\mathrm{x}}_g < x_a - \underline{\mathrm{x}}_g$. Then, in the light of Eq. \eqref{cmk3as1} and \eqref{ppk1}, we, respectively, obtain
\begin{align}
\underline{c}(\underline{\mathrm{x}}_{g},\breve{x}_b) &< \underline{c}( {\underline{\mathrm{x}}_{g},x_a}), ~~\text{if}~|\underline{\mathrm{x}}_{g} - \breve{x}_b| = |\underline{\mathrm{x}}_{g} - x_a| \label{clk1} \\
\underline{c}(\underline{\mathrm{x}}_g,0) &> \underline{c}( {\underline{\mathrm{x}}_g,x_a}), ~~\text{if}~ 0 < \underline{\mathrm{x}}_g < x_a - \underline{\mathrm{x}}_g,\label{clk2}
\end{align}
which are due to the facts: ${x}_{a} > \breve{x}_b$, $\underline{\mathrm{x}}_{g}> 0$, $|0 - \underline{\mathrm{x}}_{g}| < |\underline{\mathrm{x}}_{g} - x_{b}|$ and $x_{a} > 0$.

Considering $\breve{x}_b  < 0$, the inequalities \eqref{clk1} and \eqref{clk2} indicate that there exists a $x_b < 0$ such that $|\underline{\mathrm{x}}_{g} - x_a| > |\underline{\mathrm{x}}_{g} - x_b|$ and $\underline{c}(\underline{\mathrm{x}}_{g},x_{b}) > \underline{c}( {\underline{\mathrm{x}}_{g},{x_{a}}})$. We thus can summarize that there exists an $x_b$ such that
\begin{align}
&\underline{c}(\underline{\mathrm{x}}_{g},x_{b}) > \underline{c}( {\underline{\mathrm{x}}_{g},{x_{a}}}), ~~\text{if}~|\underline{\mathrm{x}}_{g} - x_b| < |\underline{\mathrm{x}}_{g} - x_a|, \underline{\mathrm{x}}_{g} > 0, \nonumber\\
&\hspace{5.2cm}x_b < 0, x_a > 0. \label{akm1}
\end{align}
Also, considering Eq. \eqref{cmk3as3}, if $\underline{\mathrm{x}}_g < 0$, according to the same logic used to derive Eq. \eqref{akm1}, we can conclude that there exists an $x_b$ such that
\begin{align}
&\underline{c}(\underline{\mathrm{x}}_{g},x_{b}) > \underline{c}( {\underline{\mathrm{x}}_{g},{x_{a}}}), ~~\text{if}~|\underline{\mathrm{x}}_{g} - x_b| < |\underline{\mathrm{x}}_{g} - x_a|, \underline{\mathrm{x}}_{g} < 0, \nonumber\\
&\hspace{5.3cm}x_b > 0, x_a < 0. \label{akm2}
\end{align}

Let us denote $\zeta({\underline{\mathrm{x}}}_g,x_{a},x_{b}) = \frac{|\underline{\mathrm{x}}_{g} - x_b|}{ |\underline{\mathrm{x}}_{g} - x_a|} < 1$, by which it is straightforward to verify that the union of Eq. \eqref{akm1} and \eqref{akm2} is equivalent to \eqref{confim0}. We here can conclude that the conjunctive conditions \eqref{confcc1}--\eqref{confcc1ad} result in the \underline{behavior \eqref{confim0}}.

\subsubsection*{\underline{Necessary Condition}} The necessary condition is proved via contradiction, i.e, assuming one of the conditions \eqref{confcc1}--\eqref{confcc1ad} does not hold, and then proving that the influence model $\underline{c}( {{{\underline{\mathrm{x}}}_i},{x_j}})$ cannot capture the behavior \eqref{confim3}--\eqref{confim0} simultaneously.

We assume that Eq. \eqref{confcc1} does not hold, i.e., $\underline{\mathfrak{g}}_g( {{\underline{\mathfrak{f}}_g}(\underline{\mathrm{x}}_g) - {\underline{\mathfrak{f}}_g}(x_a)})$ is nondecreasing w.r.t. the distance $| {{\underline{\mathfrak{f}}_g}(\underline{\mathrm{x}}_g) - {\underline{\mathfrak{f}}_g}(x_a)}|$. We let $\underline{\mathrm{x}}_g \geq x_{a} > 0$, and we have $| {{\underline{\mathfrak{f}}_g}(\underline{\mathrm{x}}_{g}) - {\underline{\mathfrak{f}}_g}( {{x_a}})} | = {{\underline{\mathfrak{f}}_g}(\underline{\mathrm{x}}_{g}) - {\underline{\mathfrak{f}}_g}( {{x_a}})}$. If $x_{a}$ decreases to $x_{b} > 0$, we then have $| {{\underline{\mathfrak{f}}_g}(\underline{\mathrm{x}}_{g}) - {\underline{\mathfrak{f}}_g}( {{x_a}})} | < | {{\underline{\mathfrak{f}}_g}(\underline{\mathrm{x}}_{g}) - {\underline{\mathfrak{f}}_i}( {{x_b}})} |$ and $|\underline{x}_{g} - x_{a}| < |\underline{\mathrm{x}}_{g} - x_{b}|$. As a consequence, we obtain from Eq. \eqref{conf2} that
\begin{align}
\underline{c}(\underline{\mathrm{x}}_{g},x_{a}) \leq \underline{c}( {\underline{\mathrm{x}}_{g},{x_{b}}}), \label{confim3kkl}
\end{align}
which contradicts with Eq. \eqref{confim1}. We now consider the case that $\underline{\mathfrak{g}}_g( {{\underline{\mathfrak{f}}_g}(\underline{\mathrm{x}}_g) - {\underline{\mathfrak{f}}_g}(x_a)})$ is non-decreasing w.r.t. $| {{\underline{\mathfrak{f}}_g}(\underline{\mathrm{x}}_g) - {\underline{\mathfrak{f}}_g}(x_a)}|$ and ${\underline{\mathfrak{f}}_g}(c)$ is non-increasing w.r.t. $c$. Let us set $0 < x_b < x_a < \underline{\mathrm{x}}_{g}$. We thus have $|\underline{x}_{g} - x_b| > |\underline{\mathrm{x}}_{g} - x_a|$, $|{{\underline{\mathfrak{f}}_g}( {{\underline{\mathrm{x}}_g}}) - {\underline{\mathfrak{f}}_g}( {{x_b}})} | \geq |{{\underline{\mathfrak{f}}_g}( {{\underline{\mathrm{x}}_g}}) - {\underline{\mathfrak{f}}_g}( {{x_a}})} |$ and Eq. \eqref{confim3kkl}. Following the same analysis method, we can conclude that if the conditions \eqref{confcc1}--\eqref{confcc1ad} do not hold, we  have the contradicting behavior \eqref{confim3kkl} with \eqref{confim1}.

\section*{Appendix B: Proof of Theorem \ref{tokc}}
Considering Eq. \eqref{mmc}, the second order Taylor expansion of $r(\tilde{\mathbf{u}})$ around $\mathbf{u}$ is
\begin{align}
r\left( {\tilde{\mathbf{u}} } \right) & \approx r\left( \mathbf{u}  \right) + {\left( {\tilde{\mathbf{u}}  - \mathbf{u} } \right)^\top}\frac{{\partial r}}{{\partial \mathbf{u} }} + \frac{1}{2}{\left( {\tilde{\mathbf{u}}  - \mathbf{u} } \right)^\top}\frac{{{\partial ^2}r}}{{\partial {\mathbf{u} ^2}}}\left( {\tilde{\mathbf{u}}  - \mathbf{u} } \right) \nonumber\\
& = r\left( \mathbf{u}  \right) + {\left( {\tilde{\mathbf{u}}  - \mathbf{u} } \right)^\top}\mathbf{h} + \frac{1}{2}{\left( {\tilde{\mathbf{u}}  - \mathbf{u} } \right)^\top}{\mathbf{H}}\left( {\tilde{\mathbf{u}}  - \mathbf{u} } \right). \label{tac1}
\end{align}
We now can obtain
\begin{align}
&\int\limits_{\tilde{\mathbf{u}} \in {\mathbb{U}^{\mathfrak{l}{|\mathbb{T}|}}}}\!\!\! {{e^{r\left( \mathbf{u} \right) + {{\left( {\tilde{\mathbf{u}} - \mathbf{u}} \right)}^ \top }\mathbf{h} + \frac{1}{2}{{\left( {\tilde{\mathbf{u}} - \mathbf{u}} \right)}^ \top }\mathbf{H}\left( {\tilde{\mathbf{u}} - \mathbf{u}} \right)}}} \mathrm{d}\tilde{\mathbf{u}} \nonumber\\
& = {e^{r\left( \mathbf{u} \right) + \frac{1}{2}{\mathbf{u}^ \top }\mathbf{H}\mathbf{u} - {\mathbf{u}^ \top }\mathbf{h}}}\int\limits_{\tilde{\mathbf{u}} \in {\mathbb{U}^{\mathfrak{l}{|\mathbb{T}|}}}}\!\!\!{{e^{{{\tilde{\mathbf{u}}}^ \top }\mathbf{h} - {{\tilde{\mathbf{u}}}^ \top }\mathbf{H}\mathbf{u} + \frac{1}{2}{{\tilde{\mathbf{u}}}^ \top }\mathbf{H}\tilde{\mathbf{u}}}}} \mathrm{d}\tilde{\mathbf{u}} \nonumber\\
& \approx {e^{r\left( \mathbf{u} \right) + \frac{1}{2}{\mathbf{u}^ \top }\mathbf{H}\mathbf{u} - {\mathbf{u}^ \top }\mathbf{h}}}\int\limits_{\tilde{\mathbf{u}} \in {\mathbb{U}^{\mathfrak{l}{|\mathbb{T}|}}}}\!\!\!{{e^{{{\tilde{\mathbf{u}}}^ \top }\mathbf{h} - {{\tilde{\mathbf{u}}}^ \top }\mathbf{H}\mathbf{u} }}} \mathrm{d}\tilde{\mathbf{u}} \label{pm1}\\
& = {e^{r\left( \mathbf{u} \right) + \frac{1}{2}{\mathbf{u}^ \top }\mathbf{H}\mathbf{u} - {\mathbf{u}^ \top }\mathbf{h}}}\prod\limits_{i = 1}^{\mathfrak{l}{|\mathbb{T}|}} {\int_{ - 1}^1 {{e^{{{\tilde{\mathbf{u}}}_i}{{\left[ {\mathbf{h} - \mathbf{H}\mathbf{u}} \right]}_i}}}} \mathrm{d}{{\tilde {\mathbf{u}}}_i}}, \nonumber\\
& = {e^{r\left( \mathbf{u} \right) + \frac{1}{2}{\mathbf{u}^ \top }\mathbf{H}\mathbf{u} - {\mathbf{u}^ \top }\mathbf{h}}}\prod\limits_{i = 1}^{\mathfrak{l}{|\mathbb{T}|}} {\frac{{{e^{{{\left[ {\mathbf{h} - \mathbf{H}\mathbf{u}} \right]}_i}}} - {e^{ - {{\left[ {\mathbf{h} - \mathbf{H}\mathbf{u}} \right]}_i}}}}}{{{{\left[ {\mathbf{h} - \mathbf{H}\mathbf{u}} \right]}_i}}}}, \label{pm2}
\end{align}
where  Eq. \eqref{pm1} from its previous step is obtained via considering Assumption \ref{asm1}-2).

Substituting Eq. \eqref{tac1} with Eq. \eqref{pm2} into \eqref{ahhh1zx} yields
\begin{align}
&p\left( {\mathbf{u}\left| \theta  \right.,\alpha, \tilde{x}\left( k \right)} \right) \nonumber\\
& \!\approx\! {e^{r\left( \mathbf{u} \right)}}{\left( {\int_{\tilde{\mathbf{u}} \in {\mathbb{U}^{\mathfrak{l}{|\mathbb{T}|}}}} {{e^{r\left( \mathbf{u} \right) + {{\left( {\tilde{\mathbf{u}} - \mathbf{u}} \right)}^\top}\mathbf{h} + \frac{1}{2}{{\left( {\tilde{\mathbf{u}} - \mathbf{u}} \right)}^\top}\mathbf{H}\left( {\tilde{\mathbf{u}} - \mathbf{u}} \right)}}} \mathrm{d}\tilde{\mathbf{u}}} \right)^{ - 1}} \nonumber\\
& \!=\! {e^{ - \!\frac{1}{2}{\mathbf{u}^\top}\!\mathbf{H}\mathbf{u} + {\mathbf{u}^\top}\!\mathbf{h}}}\prod\limits_{i = 1}^{\mathfrak{l}{|\mathbb{T}|}} {\frac{{{{\left[ {\mathbf{h} - \mathbf{H}\mathbf{u}} \right]}_i}}}{{{e^{{{\left[ {\mathbf{h} - \mathbf{H}\mathbf{u}} \right]}_i}}} - {e^{ - {{\left[ {\mathbf{h} - \mathbf{H}\mathbf{u}} \right]}_i}}}}}},\nonumber
\end{align}
by which we obtain $\widetilde{\mathcal{L}}\left( {\theta ,\alpha } \right) \approx \log p\left( {\mathbf{u}\left| \theta  \right.,\alpha, \tilde{x}(k)} \right)$, where
$\widetilde{\mathcal{L}}\left( {\theta ,\alpha } \right)$ is given in Eq. \eqref{ahhhvg}.
\bibliographystyle{IEEEtran}
\bibliography{ref}
\end{document}